\documentclass{aa}
\usepackage{graphics,latexsym,amssymb,times,psfig}

\begin{document}

\title{Time Resolved Spectral Analysis of Bright Gamma Ray Bursts}

\author{G. Ghirlanda\inst{1}\and A. Celotti\inst{1}
\and G. Ghisellini \inst{2}}

\offprints{G. Ghirlanda; email: ghirland@sissa.it}
\institute{SISSA/ISAS, Trieste - V. Beirut 2-4, 34014\\
\and    Osservatorio Astronomico di Brera, via Bianchi 46, Merate (LC), Italy}
\date{21 June 2002}
\titlerunning{Spectra of Bright Gamma Ray Bursts}
\authorrunning{G. Ghirlanda, A. Celotti \& G. Ghisellini}

\abstract{ We present the time
integrated and time resolved spectral analysis of a sample
of bright bursts selected with $F_{peak} \ge 20\
    \mathrm{phot\ cm^{-2}\ sec^{-1}}$ from the BATSE archive.  We fitted four
    different spectral models to the pulse time integrated and time
    resolved spectra. We compare the low energy slope of the fitted spectra
    with the prediction of the synchrotron theory
    [predicting photon spectra softer than $N(E)\propto E^{-2/3}$],
    and test, through direct spectral fitting, the synchrotron shock model.
    We point out that differences in the parameters distribution can be
    ascribed to the different spectral shape of the models employed
    and that in most cases the spectrum can be described by a smoothly
    curved function. The synchrotron shock model
    does not give satisfactory fits to the time averaged and time
    resolved spectra. Finally, we derive that the synchrotron low
    energy limit is violated in a considerable number of spectra both
    during the rise and decay phase around the peak.
 }
\maketitle

\section{Introduction}

The nature and emission mechanisms responsible for the
prompt emission of Gamma--ray bursts (GRB) are still a
matter of debate. From the phenomenological perspective
much effort has been made in order to
characterize and identify typical spectral properties of
bursts, mostly by applying parametric but very general and
simple spectral models. Probably the most widely adopted is
that suggested by Band et al. (\cite{Band b}), namely a
smoothly connected double power law model. Its attractive
feature is that it characterizes, within the
observational energy window, the most relevant quantities,
namely the peak energy, representing the energy at which
most of the emission occurs, and the low and high energy
components which are related, according to the most
accredited emission theories, to the particle energy
distribution and/or to the physical parameters of the
emitting region.
Indeed GRB spectra result to be well represented
by the Band parameterization ($N(E)\propto E^{\alpha , \beta}$),
with typical low energy power law photon
spectral indices $\alpha$ between --1.25 and --0.25,
high energy spectral indices
$\beta$ around --2.25 (Preece et al. \cite{Preece1998}) and
peak energy $E_{p}$ typically around 100--300 keV.

A fundamental  issue in the  spectral analysis of GRBs  is the
integration timescale of  the spectra which are observed  to vary even
on  millisecond timescales  (Fishman et  al. \cite{Fishman  b}). BATSE
spectra,  which  are integrated  for  a  minimum  of 128ms,  therefore
represent  the best  way, presently available, to  constrain the  burst  emission mechanism. Furthermore,  in  order to  compare the  average
spectral properties of different bursts, also time integrated spectra,
covering the duration of the pulse or the entire burst, have been used
in literature.

It has been found in general that the Band model not only
well describes the time integrated spectra, but it also
appears to fit the time resolved spectra of bright bursts.
Alternative spectral models -- and in particular the
predictions for synchrotron emission (Katz \cite{Katz},
Crider at al. \cite{Crider a},b) -- have also been recently
tested on a large sample of time resolved spectra by Preece
and collaborators (Preece et al. \cite{Preece2000}).

Within  this scenario,  here  we  present the  study  of the  spectral
properties  of  single pulses  within  bright  GRBs  which intends  to
complement  the work mentioned  above, by  specifically aiming  at: 1)
compare the  results of the  analysis of the  spectra averaged
over  major pulses  in the  burst lightcurve  with the  time resolved
spectra  of  the very  same  burst  in  order to  quantify  systematic
differences; 2) consider both empirical (Band model) and more
`physical' (synchrotron shock  model) spectral models for all
spectra  and  compare  the  quality  of  the  corresponding  fits.  In
particular  each spectrum  is fitted  with the  four models  we choose
(Band's,  broken power--law,  thermal  Comptonization and  synchrotron
shock model, the latter fitted  to temporal resolved BATSE spectra for
the  first  time). We  also  examine  any  spectral `violation'  (with
respect to  the predicted slope  in the case of  synchrotron emission)
for the entire  burst evolution.  The development of  this work is the
analysis and interpretation of  the spectral evolution morphologies of
this sample of bursts.

The paper is structured as follows: In sect. 2 we
describe the data and their selection criteria, while the
spectral models adopted for the analysis are detailed in
sect. 3. Section 4 presents the results of our work for
the time integrated and time resolved spectra.
Conclusions are drawn in sect. 5.  The time
evolution of the spectral parameter will be the content
presented in a following paper (Ghirlanda et al. in
preparation).

\section{Data selection and analysis}

\subsection{Instrumental summary}

The Burst and Transient Source Experiment (BATSE) consisted
of eight detection modules on the corners of the Compton
Gamma Ray Observatory (CGRO -- deorbited in summer 2001). 
Each module was composed by two instruments: the Large Area 
Detector (LAD) designed for burst location and temporal analysis, 
and the Spectroscopic Detector (SD) suited for spectral study.

Each LAD consisted of a thin circular \emph{NaI} layer of 2025 $cm^{2}$ collecting area.
The nominal energy coverage was 28 -- 1800 keV (with minor
variations between different detector modules). The energy
resolution was about 20\% at 511 keV (Fishman et
al. \cite{Fishman a}).

The SDs were smaller (127 $cm^{2}$) but thicker
and thus had a greater energy conversion efficiency.
Their sensitive energy range varied with the gain settings
and energy thresholds of the photomultipliers, but
typically extended from 20 keV to 2 MeV (Band et al.
\cite{Band a}). The energy resolution is 7\% at 662 keV
(Fishman et al. \cite{Fishman a}, Band et al. \cite{Band b}).

After a burst trigger, during the following 4 minutes burst
mode, the LADs accumulated, among other data products, the
\emph{High Energy Resolution Burst} (HERB) spectra. These
data are a sequence of 128 quasi--logarithmically spaced energy
channel spectra with a maximum time resolution of 128 ms. 
They were accumulated from the 4 most illuminated detectors.
The SDs produced similar spectra with 256 energy channels
and 128 ms resolution (SHERB). The maximum number of
spectra accumulated is 128 and 192 for the HERB and SHERB
data, respectively.
The LAD and SD spectral time integration algorithm was 
based on a count rate criterion so that the instruments provided spectra with a minimum integration time of 128 ms only for particularly intense bursts or around the peaks, while in most cases the spectral accumulation timescale was much greater than 128 ms.

\subsection{Data Selection}       \label{sez:dataselection}

The HERB data have been systematically
preferred for this work because the higher detection area of the LAD
ensures an higher count rate than the SD detectors. Despite
their moderate energy resolution (if compared to the SDs)
they are suited for the continuum spectral study (Preece et
al. \cite{Preece2000}).

There are a few cases (see col. 3 in Table \ref{GRBsamp})
for which the SHERB data were used because there were
telemetry gaps.

We have analyzed the HERB data from rank 1 detector.
The rank of the detector (col. 4 in Table \ref{GRBsamp})
is an indication of the relative count rate during the
trigger: rank 1 detector has the highest count rate, the
best S/N and the highest spectral time resolution. For the SHERB
data the detector choice is a compromise between the
highest degree of illumination (the first rank SD detector)
and the highest gain which depends on different ground
setting parameters, which have been changed during the
mission (Band D., private communication). The gain of the
photomultiplier, in fact, scales the spectrum up or down in
energy: the highest is the gain the more the detector is
sensible to the low energy photons and the lower is the low
energy threshold (Preece et al., \cite{Preece b}; Kitchin
\cite{Kitchin}). We selected the highest rank SD detector
with the 511 keV calibration line centroid above uncompressed channel 800.

\subsection{The bursts sample}

The bursts were selected from the BATSE 4B
catalog\footnote{http://cossc.gsfc.nasa.gov/batse/4Bcatalog/index.html.}
which is complete until 29 Aug 1996 (trigger 5586). This
catalog was complemented with the on line catalog which
includes the triggers after 5586 until trigger 8121 (9 Sept
2000).

We have selected GRBs with a peak flux on the 64 ms time
scale (calculated according to Fishman et al.
\cite{Fishman1}, Meegan et al. \cite{Meegan1}, Paciesas et
al. \cite{Paciesas}) higher than 20 $\mathrm{phot\ cm^{-2}\ sec^{-1}}$.
This choice is motivated by the fact that bright
bursts should provide time resolved spectra with good S/N
(also on integration of the order of 128 ms).
We collected in this way 38 bursts whose trigger
number, name and peak flux are reported in
Table.~\ref{GRBsamp} (cols. 1, 2 and 6, respectively).

The sample selected according to the peak flux criterion
was reduced during the spectral analysis because of data
problems (trigger 1609, 1711, 3480) or because the number
of spectra available for the
spectral analysis was less than 5 (trigger 1997, 2151,
2431, 2611, 3412, 5711, 5989, 6293, 6904, 7647).
This happened particularly in short bursts which have
typical duration lower than 1 second. The bursts which were not analyzed have dashes entries in tab~\ref{GRBsamp}. The
final sample of analyzed bursts contains 25 bursts and is obviously not complete with respect to the flux selection criterion.
All the analyzed bursts except trigger 5563,7301,7549 are also 
present in Preece et
al. (\cite{Preece2000}) spectral sample.
 \begin{table*}
      \caption[]{The GRB sample}
         \label{GRBsamp}
      \[ { \scriptsize
         \begin{array}{ccccccccccc}
                \hline
            \noalign{\smallskip}
   Trigger $$^{\mathrm{a}}$$ & GRB & Data  & \multicolumn{2}{c}{Detector} & Peak Flux $$^{\mathrm{c}}$$ & \multicolumn{2}{c}{Background} &  Peak$$^{\mathrm{e}}$$ &
Spectra$$^{\mathrm{f}}$$  & S/N$$^{\mathrm{g}}$$ \\
&                    & Type $$^{\mathrm{b}}$$  & Number
&  Rank     & [phot/cm^2/sec] & \#$$^{\mathrm{d}}$$ & n
$$^{\mathrm{d}}$$  &     &         & $$\ge$$  \\
\noalign{\smallskip}
                \hline
            \noalign{\smallskip}
      143   & \object{910503}  & HERB  & LAD6  &  1  & $$52.08\pm 1.43$$ & 8  & 3 & 4 & 49 & - \\
      1473  & \object{920311}  & HERB  & LAD5  &  1  & $$25.31\pm 0.7$$  & 6  & 2 & 3 & 55 & - \\
      1541  & \object{920406}  & SHERB & SD2   &  2  & $$38.32\pm 0.89$$ & 5  & 2 & 1 & 9  & 15\\
      1609  & 920517  &  --   &  --   &  -  & $$67.59\pm 1.21$$ & -  & - & - & -  & - \\
      1625  & \object{920525}  & HERB  & LAD4  &  1  & $$28.06\pm 0.74$$ & 8  & 4 & 3 & 25 & - \\
      1711  & 920720  &  --   &  --   &  -  & $$21.7\pm 0.7$$   & -  & - & - & -  & - \\
      1997  & 921022  &  --   &  --   &  -  & $$40.33\pm 0.85$$ & -  & - & - & -  & - \\
      2083  & \object{921207}  & HERB  & LAD0  &  1  & $$46.55\pm 0.92$$ & 6  & 2 & 2 & 31 & - \\
      2151  & 930131  &  --   &  --   &  -  & $$167.84\pm 2.63$$& -  & - & - & -  & - \\
      2329  & \object{930506}  & SHERB  & SD2  &  4  & $$42.57\pm 0.90$$ & 11 & 4 & 1 & 42 & - \\
      2431  & 930706  &  --   &  --   &  -  & $$43.83\pm 0.89$$ & -  & - & - & -  & - \\
      2537  & \object{930922}  & HERB  & LAD1  &  1  & $$27.28\pm 0.7$$  & 31 & 4 & 3 & 23 & 45\\
      2611  & 931031  &  --   &  --   &  -  & $$35.05\pm 0.89$$ &    &   &   &    &   \\
      2798  & \object{940206}  & HERB  & LAD3  &  1  & $$24.19\pm 0.73$$ & 7  & 3 & 1 & 16 & - \\
      2831  & \object{940217}  & HERB  & LAD0  &  1  & $$44.34\pm 1.08$$ & 10 & 4 & 3 & 28 & - \\
      3412  & 950211  &  --   &  --   &  -  & $$54.82\pm 0.76$$ & -  & - & - &  - & - \\
      3480  & 950325  &  --   &  --   &  -  & $$21.61\pm 0.51$$ & -  & - & - & -  & - \\
      3481  & \object{950325}  & HERB  & LAD2  &  1  & $$25.7\pm 0.6$$   & 6  & 4 & 3 & 27 & - \\
      3491  & \object{950403}  & HERB  & LAD3  &  1  & $$30.65\pm 0.62$$ & 6  & 3 & 3 & 44 & 45\\
      3492  & \object{950403}  & HERB  & LAD5  &  1  & $$61.44\pm 0.91$$ & 6  & 4 & 1 & 22 & 45\\
      3523  & \object{950425}  & HERB  & LAD6  &  1  & $$21.81\pm 0.53$$ & 4  & 2 & 2 & 50 & - \\
      4368  & \object{960114}  & SHERB  & SD0  &  1  & $$58.61\pm 0.83$$ & 8  & 4 & 3 & 29 & 45\\
      5477  & \object{960529}  & HERB  & LAD1  &  1  & $$29.35\pm 0.70$$ & 7  & 4 & 2 & 12 & - \\
      5563  & \object{960804}  & HERB  & LAD4  &  1  & $$22.7\pm 0.59$$  & 7  & 4 & 1 & 8  & - \\
      5567  & \object{960807}  & HERB  & LAD0  &  1  & $$22.8\pm 0.6$$   & 6  & 4 & 5 & 25 & 45\\
      5614  & \object{960924}  & SHERB  & SD6  &  1  & $$183.37\pm 1.62$$& 12 & 4 & 1 & 22 & - \\
      5621  & \object{961001}  & HERB  & LAD2  &  1  & $$25.64\pm 0.61$$ & 31 & 4 & 2 & 17 & - \\
      5704  & \object{961202}  & HERB  & LAD0  &  1  & $$43.93\pm 0.76$$ & 26 & 4 & 1 & 13 & - \\
      5711  & 961212  &  --   &  --   &  -  & $$41.25\pm 0.71$$ & -  & - & - & -  & - \\
      5989  & 970201  &  --   &  --   &  -  & $$77.61\pm 0.89$$ & -  & - & - & -  & - \\
      6198  & \object{970420}  & HERB  & LAD4  &  1  & $$66.5\pm 0.9$$   & 12 & 4 & 1 & 38 & - \\
      6293  & 970704  &  --   &  --   &  -  & $$88.53\pm 1.$$   & -  & - & - & -  & - \\
      6404  & \object{970930}  & HERB  & LAD6  &  1  & $$24.01\pm 0.52$$ & 21 & 4 & 2 & 22 & - \\
      6581  & \object{980125}  & HERB  & LAD0  &  1  & $$40.91\pm 0.69$$ & 15 & 4 & 1 & 17 & - \\
      6904  & 980706  &  --   &  --   &  -  & $$54.62\pm 0.87$$ & -  & - & - & -  & - \\
      7301  & \object{990104}  & HERB  & LAD7  &  1  & $$86.53\pm 0.93$$ & 12 & 4 & 5 & 24 & - \\
      7549  & \object{990506}  & HERB  & LAD7  &  1  & $$25.12\pm 0.58$$ & 9  & 4 & 6 & 52 & - \\
      7647  & 990712  &  --   &  --   &  -  & $$24.06\pm 0.51$$ & -  & - & - & -  & - \\
        \noalign{\smallskip}
        \hline
    \end{array}
      } \]
\begin{list}{}{}
\item[$^{\mathrm{a}}$] Burst reference number from the Gamma Burst Catalog at \\
         $http://gammaray.msfc.nasa.gov/batse/grb/catalog/current/$
\item[$^{\mathrm{b}}$] (S)HERB: (Spectroscopic) High Energy Resolution Burst data.
\item[$^{\mathrm{c}}$] Peak flux on the 64 ms time-scale, integrated over the 50 - 300 keV energy range.
\item[$^{\mathrm{d}}$] Background is calculated on '' \# '' number of spectra and with a polynomial function of degree ''n''.
\item[$^{\mathrm{e}}$] Number of pulses spectroscopically analyzed within each burst.
\item[$^{\mathrm{f}}$] Number of time resolved spectra per burst.
\item[$^{\mathrm{g}}$] S/N method used in grouping the  time resolved spectra accumulated by the instrument.
\end{list}
   \end{table*}
%

%
%
\begin{figure}
\resizebox{\hsize}{!}{\includegraphics{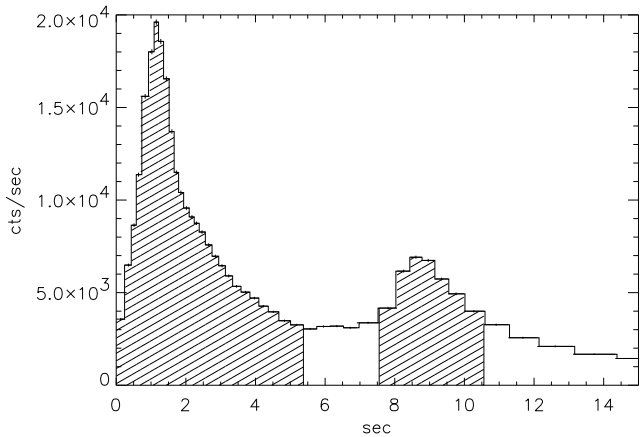}}
\caption[]{Example of light curve (trigger 2083).
      Each time bin represents the
      count rate summed over all the detectors that triggered, integrated in energy from 28 to 1800 keV and over the detection area
      (for details see Preece et
      al. \cite{Preece2000}).  The line--filled regions show the time
      intervals of the average pulse spectrum accumulation.  The start
      and stop times were chosen to have the same flux level.
      Our analysis is limited to the dashed areas, and therefore
      excludes the long slowly decaying part of the burst.}
      \label{fig1} \end{figure}

\section{Spectral analysis}         \label{sez:spectralana}

The spectral analysis has been performed with the software
\emph{SOAR} v3.0 (Spectroscopic Oriented Analysis Routines)
by Ford (\cite{Ford}).
The power of this software is its multi--level routine
structure which can be easily modified by
the user: we implemented the spectral library with the
broken power law and added the possibility of fitting each time
resolved spectrum with a different energy binning scheme.
The main capabilities of SOAR are: the possibility of
analyzing HERB and SHERB data and rebinning (in time) spectra
in different fashions (according to a fixed number of
spectra per group or with a S/N ratio
criterion). It is also possible to simulate burst spectra
and spectral evolution (Ford \cite{Ford}).

We used the package XSPEC, as widely used for
high energy data,  to test,  for two bursts, the spectral analysis
procedure. We found, within parameters errors, similar results. Any difference
can be ascribed to a different background calculation technique.

The spectral analysis procedure consisted in the inspection
of the light curve, i.e. the count rate in cts s$^{-1}$
over the nominal energy band (typically 28--1800 keV) with
each time bin corresponding to a time resolved spectrum
(with typical integration time of 128 ms).

The background is calculated on a time interval as close as
possible to the burst, but not contaminated by the burst
itself. Typically we use a $1000\ \mathrm{sec}$ interval
before and after the GRB, and fit the spectra contained in
this interval with a polynomial background model examining
the  residuals  for any systematic problem.  The background
spectrum to be subtracted is obtained by extrapolating the
polynomial best fit spectrum, for every channel, in the
time interval occupied by the burst.  
As reported in Table~\ref{GRBsamp}, in most cases the degree of the
polynomial is 4.

The GRB light  curve was used to identify  the pulses as those
structures  in which  the  flux rises  above  and decays  back to  the
background level. We accumulated the time resolved spectra within this
time interval: this defines the  time average pulse spectrum. In those
pulses composed  by several  substructures we treated  them separately
only if their  flux starts and returns to the  background level and if
they are separated by at least a 128ms time bin.
In the example trigger of Fig.~\ref{fig1} we accumulated 
two average pulse spectra,
corresponding to the dashed regions in the plot. The time
averaged pulse spectrum was then fitted with the 4 spectral
models described in the following section. The best fit
results of the average spectrum were used as initial
parameters guess for fitting the time resolved spectra.

For the time resolved spectra we used, as integration time,
at first the shortest time available (i.e. 128 ms, at best),
again limiting our analysis to the same time--span selected
for the time integrated analysis. The best fit parameters
were then examined in search for any indetermination: when
at least 3 subsequent spectra had one or more undetermined
spectral parameter, the spectra were grouped according to a
S/N criterion.
We followed the prescriptions of Preece et al. (\cite{Preece2000}) who
accumulate subsequent HERB spectra until the S/N
(calculated for the 28--1800 keV energy range) is greater
than 45; for the SHERB data the S/N threshold was fixed at 15 as
already done by Ford et al. (\cite{Ford a}). If after this
accumulation the number of time resolved spectra was less
than 5 the burst was not included in the final list of
analyzed events (see sect.\ref{sez:dataselection}).

The spectra were also rebinned in energy in order to be
confident that the Poisson statistic is represented by a
normal distribution in every channel for the application of
the $\chi^2$ minimization technique.  We fixed the minimum
number of counts per energy bin at 30 and 15 for the HERB
and SHERB data, respectively.

\subsection{Spectral Models}     \label{sez:spectralmodel}

We used four spectral representations to fit the GRB time
averaged and time resolved spectra. These functions were
chosen in order to find whether a specific model can be
considered as best representation of the spectral characteristics
of pulses in bright GRBs. We explicitly exclude a fifth
spectral model proposed in the literature by Ryde
(\cite{Ryde}) because it has a higher number of free
parameters.

\subsubsection{The Band model}

This empirical model (BAND hereafter) was first proposed by
Band et al. (\cite{Band b}) and, as already mentioned,
is a good general and simple description of the time averaged (Band et
al.\cite{Band b}) and the time resolved spectra (Ford et
al \cite{Ford a}; Preece et al. \cite{Preece1998}). It
contains the two continuum components in the keV--MeV band
already discovered before BATSE: (i) a low energy power law
with an exponential cutoff $N(E)\propto E^{\alpha}
exp(-E/E_{0})$
and (ii) a high energy power law $N(E)\propto E^{\beta}$
(Matz et al. \cite{Matz}). In fact the BAND model
consists of 2 power laws joined smoothly by an exponential
roll--over, namely:
\begin{equation}
N(E)=
\left\{
\begin{array}{l}
A \left(\frac{E}{100\ \mathrm{keV}}\right)^\alpha \exp \left( - {E \over E_0} \right)\\
\hspace{105pt} \mathrm{for}\ E \leq \left( \alpha - \beta \right) E_{0}  \\
A E^{\beta}\left[\frac{(\alpha-\beta)E_0}{100\ \mathrm{keV}}\right]^{\alpha-\beta} \exp(\beta-\alpha)\\
\hspace{105pt}  \mathrm{for}\ E \geq \left( \alpha - \beta \right) E_{0} 
\end{array}\right.
\end{equation}
where $N(E)$ is in photons cm$^{-2}$ s$^{-1}$ keV$^{-1}$. The
free parameters, which are the result of the fits, are:
\begin{itemize}
     \item A: the normalization constant @ 100 keV;
     \item $\alpha$: the low energy power law spectral index;
     \item $\beta$: the high energy power law spectral index;
     \item $E_{0}$: the break energy, which represents the
              e--folding parameter.
\end{itemize}
If $\beta< -2$ the peak energy in the $E\ F_E$ diagram ($F_E$
is the flux in keV/$\mathrm{cm^{-2} sec^{-1}}$ keV) is
$E_{peak}=(\alpha + 2)E_{0}$ and represents the energy at
which most of the power is emitted.

For the fitting procedure we had to assume an interval for
the parameters and we fixed [$-5$,1]
for $\alpha$ and $\beta$ and the break energy was allowed
to vary in the range
[28--1800] keV.

\subsubsection{The Broken Power Law model}

This model, called BPLW, is the simplest model used for
fitting GRB spectra: it consists of two power laws sharply
connected, with no curvature. Its analytical form is:

\begin{equation}
N(E)=\left\{
\begin{array}{lll}
A E^\alpha\ \ & \mathrm{for} & E \leq\left( \alpha - \beta \right) E_{0}\\ 
A E_{0}^{\alpha - \beta} E^\beta\ \ & \mathrm{for} & E \geq\left( \alpha - \beta \right) E_{0}
\end{array}\right.
\end{equation}
The free parameters are as before. In this model the peak
energy of the $E F_E$ diagram coincides with the break
energy $E_{0}$ for $\beta < -2$.

\subsubsection{The Comptonization model}

This spectral representation (COMP hereafter) is composed
of a power law ending-up in an exponential cutoff, thus fitting
well those spectra with a very steep high energy decline:
\begin{equation}
     N(E)=A E^{\alpha} \exp
     \left( -\frac{E}{E_{0}}\right).
\end{equation}
The free parameters are as in the previous models without
the high energy component. In fact the COMP model can be 
analytically obtained from the BAND model assuming $\beta=-\infty$. 
This model might be considered to mimic the spectrum resulting 
from multiple Compton emission from a thermal medium.

\subsubsection{The Synchrotron Shock Model}

The model (SSM) is based on optically thin synchrotron
emission from relativistic particles (either electrons and/or
electron--positron pairs) (Tavani \cite{Tavani b}). 
The electron energy distribution $N(\gamma)$, which is a relativistic Maxwellian before the shock occurs, is modeled for the internal shock passage by adding a high energy powerlaw component (Tavani \cite{Tavani b}): 
\begin{equation} N\left( \gamma \right) \propto
\left\{\begin{array}{lll}
\frac{\gamma^{2}}{\gamma_{\star}^{3}} \exp \left( -\frac{\gamma}{\gamma_{\star}} \right)\ \ \ & \mathrm{for} & \frac{\gamma}{\gamma_{\star}} \leq 1 \label{postshock} \\
\gamma^{-\delta}\ \ \ & \mathrm{for} & 1 \leq \frac{\gamma}{\gamma_{\star}} \leq  \frac{\gamma_{max}}{\gamma_{\star}} 
\end{array}\right.
\end{equation}
where $\gamma m_{e} c^{2}$ is the electron energy,
$\gamma_{\star} = k_{B} T^{\star}/m_{e} c^{2}$ is the
pre-shock equilibrium electron energy, $\gamma^{-\delta}$
is the supra-thermal component and $\gamma_{max}$ is the
maximum electron energy.
The power law part of $N(\gamma)$ (of index $\delta$)
is related to the high energy spectral power law component
of the photon spectrum ($\propto E^{\beta}$) by the relation
$\beta=-(\delta+1)/2$.


   \begin{figure}
    \resizebox{\hsize}{!}{\includegraphics{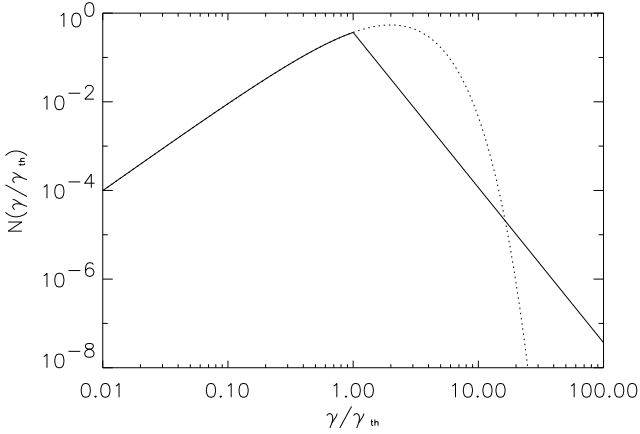}}
      \caption[]{Different electron energy distributions,
      \textit{solid line}=post shock Maxwellian with a high energy
      power law tail ($\delta = 3.5$); \textit{dotted line}=pre-shock
      relativistic Maxwellian. The distributions are functions of
      $\gamma/\gamma{\star}$, and $N(\gamma /\gamma{\star})$ is in
      arbitrary units. } \label{fig2} \end{figure}
%

%
   \begin{figure}
    \resizebox{\hsize}{!}{\includegraphics{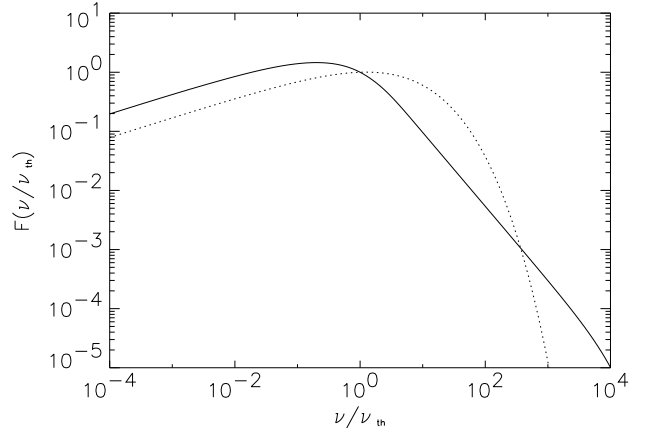}}
      \caption[]{Synchrotron spectra corresponding to the two
      electron energy distributions of Fig.~\ref{fig2}: \textit{solid
      line}=post shock Maxwellian with a high energy power law tail
      ($\delta = 3.5$); \textit{dotted line}=pre-shock relativistic
      Maxwellian. The distributions are functions of $\nu/\nu{\star}$.
      The shown spectra are normalized to their respective peak flux.}
 \label{fig3} \end{figure}

In Fig.~\ref{fig2} we report two cases of electron energy
distribution. Notice that in the case described by 
Eq.~\ref{postshock} (solid line), the distribution
extends on a wider energy range compared with the
relativistic Maxwellian. In the figure a
typical intermediate value for the power law index
$\delta=3.5$ is assumed.

The synchrotron spectrum emitted by such an electron population can be obtained via the classic formul\ae  (Rybicki \& Lightman \cite{Rybicki}):
\begin{eqnarray}
I_{syn} \propto  B_{ps} & ( & \int_{0}^{1}x^{2}exp(-x)\ F_{5/3}(\frac{y}{x^{2}})\ dx\ +\nonumber\\
                        &         &+ \frac{1}{e}\int_{1}^{x_{max}}x^{-\delta}\ F_{5/3}(\frac{y}{x^{2}})\ dx\ )\nonumber
\end{eqnarray}

where $x=\gamma/\gamma_{\star}$, $y=E/E_{ch}^{\star}$ and $E_{ch}^{\star}$ is the synchrotron characteristic energy corresponding to those electrons with energy $E_{\star}=m_{e}c^{2}\gamma_{\star}$.
$B_{ps}$ is the post--shock magnetic field which is the pre-existing B enhanced by the strong shock passage (Kennel et al \cite{Kennel}).

The shape of the synchrotron spectrum emitted by this electron
population (Fig. \ref{fig3}, solid line) has a substantial
continuous curvature and is characterized by a low energy
component (namely for $E<E_{peak}\sim 2.5 E_{\star}$, where $E_{\star}=m_{e}c^{2}\gamma_{\star}$ is the electron energy) which should be well represented by the single electron
synchrotron spectrum dependence $\sim E^{4/3}$ (in the $E
F_E$ diagram, i.e. $\sim E^{-2/3}$ in the count spectrum)
and by $E^{-(\delta - 3)/2}$ at high energies (i.e. above
$E_{peak}$). It is evident that this spectral form can
account for the high energy spectral variety of GRB
observed spectra but has a fixed ($-2/3$) slope at low
energies.
One immediate prediction of this model is the relation between the electron energy $\gamma_{\star}$, the post--shock magnetic field $B_{ps}$ and the photon spectrum peak energy $E_{ch}^{\star}=\left(3\ e\ h / 4\pi\ m_{e}\right)\ B_{ps}\gamma_{\star}^{2}$, which implies that the product $<B_{ps}\gamma_{\star}^{2}>$ can be constrained by the data.

The free parameters of the model are the high energy
electron spectral index $\delta$ and the characteristic energy
$E_{\star}$ which describe the electron energy distribution. 
In the fit of this model to the time resolved
spectra we assumed that the maximal interval of variation
of these 2 parameters is $\delta \in [1, 10]$ and
$\mathrm{E_{\star}} \in [10-2000] \mathrm{keV}$.

\subsubsection{Comparison of the fitting functions}

\begin{figure}
    \resizebox{\hsize}{!}{\includegraphics{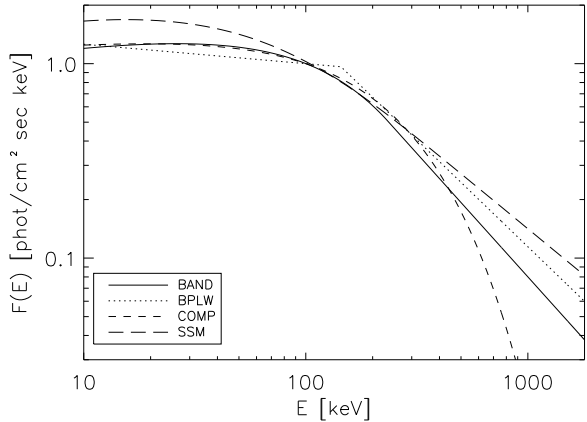}}
      \caption[]{The 4 spectral models: the spectral parameters correspond to the average values 
of the pulse time integrated spectra reported in Table~\ref{averagefits}.}           \label{4modelli}
   \end{figure}

For clarity, in Fig.~\ref{4modelli} we report the spectral
shapes corresponding to the four models considered.  The
potential of each model to better represent a particular
spectral shape is clearly visible. 
The four models are similar at low energies but they
differ in the high energy tail. In general we expect the
average COMP model to overestimate the
spectral break energy due to the lack of a 
power law high energy component.
Furthermore the average BAND spectrum would tend to be harder 
at low energies and softer at high energies than the BPLW model, due to the
sharp break of the latter. The SSM average spectrum results
harder than the BAND and COMP model as the low energy
spectral index is fixed in the SSM model, making
the average spectral shape harder. 

We tested all the 4 spectral models presented above in order to be consistent with the previous work present in literature (e.g. Preece et al. \cite{Preece2000}). We notice, anyway, that the BPLW model is unphysical due to its sharp spectral break and the COMP model can be considered a subset of the more general BAND form, leaving the latter and the SSM as the primary spectral models to be compared.

\section{Results}

We have fitted the 25 GRBs of the sample with the four
spectral models presented in sect.~\ref{sez:spectralmodel}.
For each peak present in the burst light curve we have
analyzed the time integrated spectrum and the sequence of
time resolved spectra. As mentioned in the Introduction the
goals are the following:

$\bullet$ determine whether there is a preferential shape
which better represents the time integrated and/or the time
resolved spectra of bright bursts;

$\bullet$ alternatively, to determine the robustness of the
spectral quantities with respect to the specific model
considered;

$\bullet$ quantify the difference between time integrated
and time resolved spectra of the very same peak;

$\bullet$ verify the strongest prediction of the
synchrotron theory of a low energy spectral shape not
evolving and fixed to $-$2/3 (Katz et al., \cite{Katz}).

For clarity we first present the results for the time
integrated (sect. 4.1), then for the time resolved
(sect. 4.2) spectra, and finally discuss the violation of
the synchrotron 'limit' (sect. 4.3).


\subsection{Time integrated pulse spectrum}

\begin{figure*}
\centerline{\psfig{file=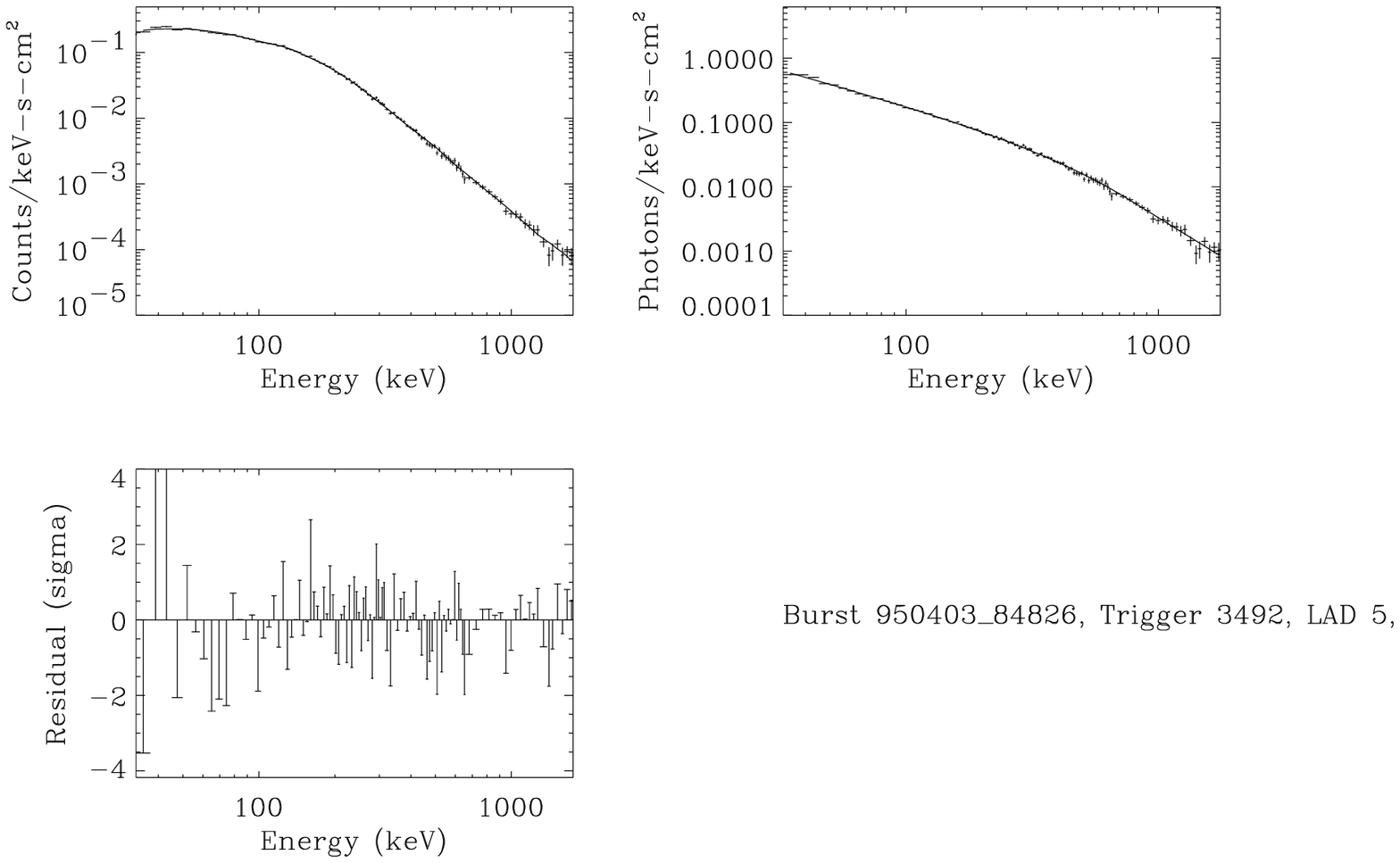,bbllx=62pt,bblly=578pt,bburx=295
pt,bbury=704pt,clip=true}\psfig{file=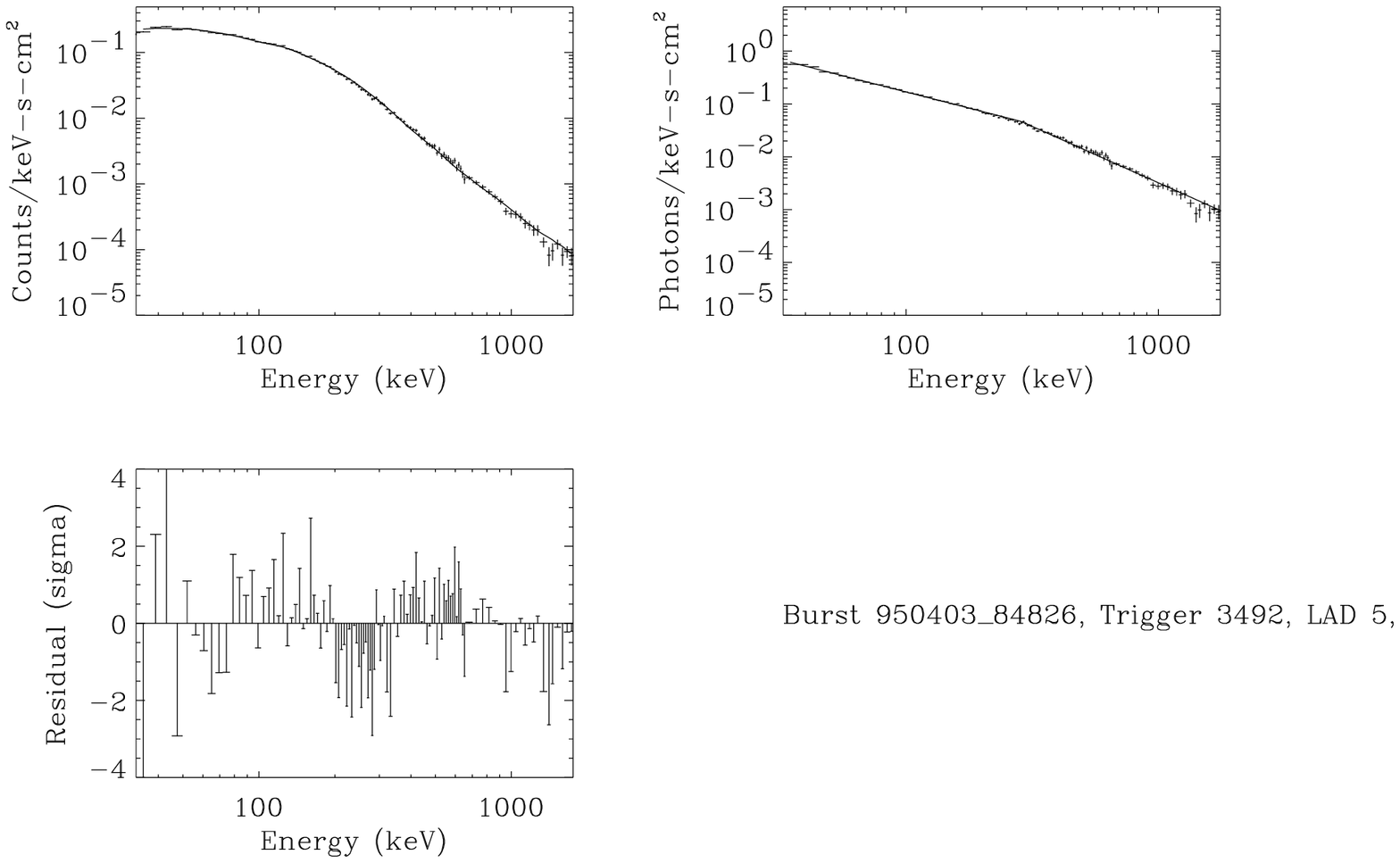,bbllx=116pt,bblly=578pt,bburx=295
pt,bbury=704pt,clip=true}}
\centerline{\psfig{file=grb.ps,bbllx=62pt,bblly=368pt,bburx=295
pt,bbury=524pt,clip=true}\psfig{file=bplw.ps,bbllx=116pt,bblly=368pt,bburx=295
pt,bbury=524pt,clip=true}}
\centerline{\psfig{file=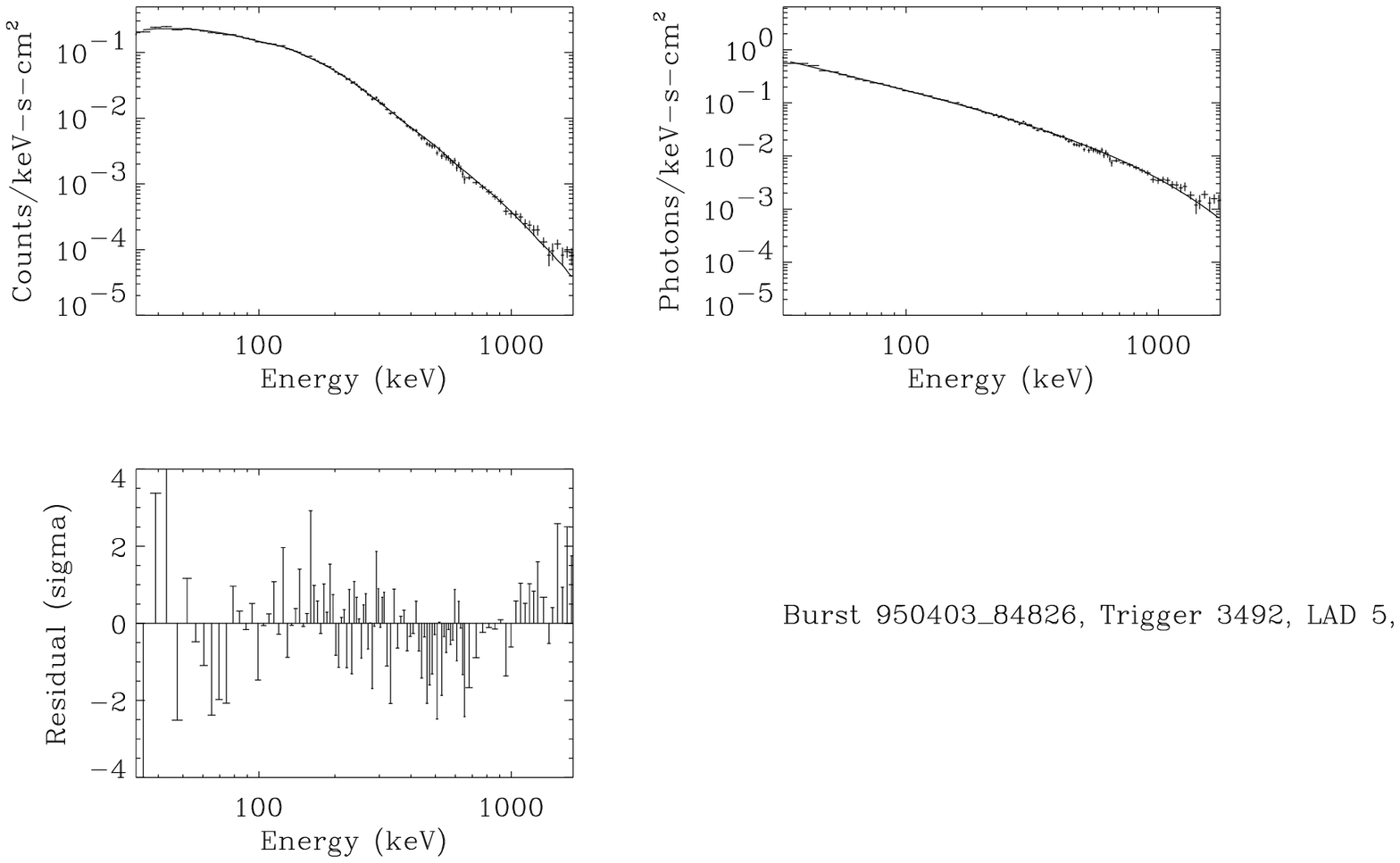,bbllx=62pt,bblly=578pt,bburx=295
pt,bbury=704pt,clip=true}\psfig{file=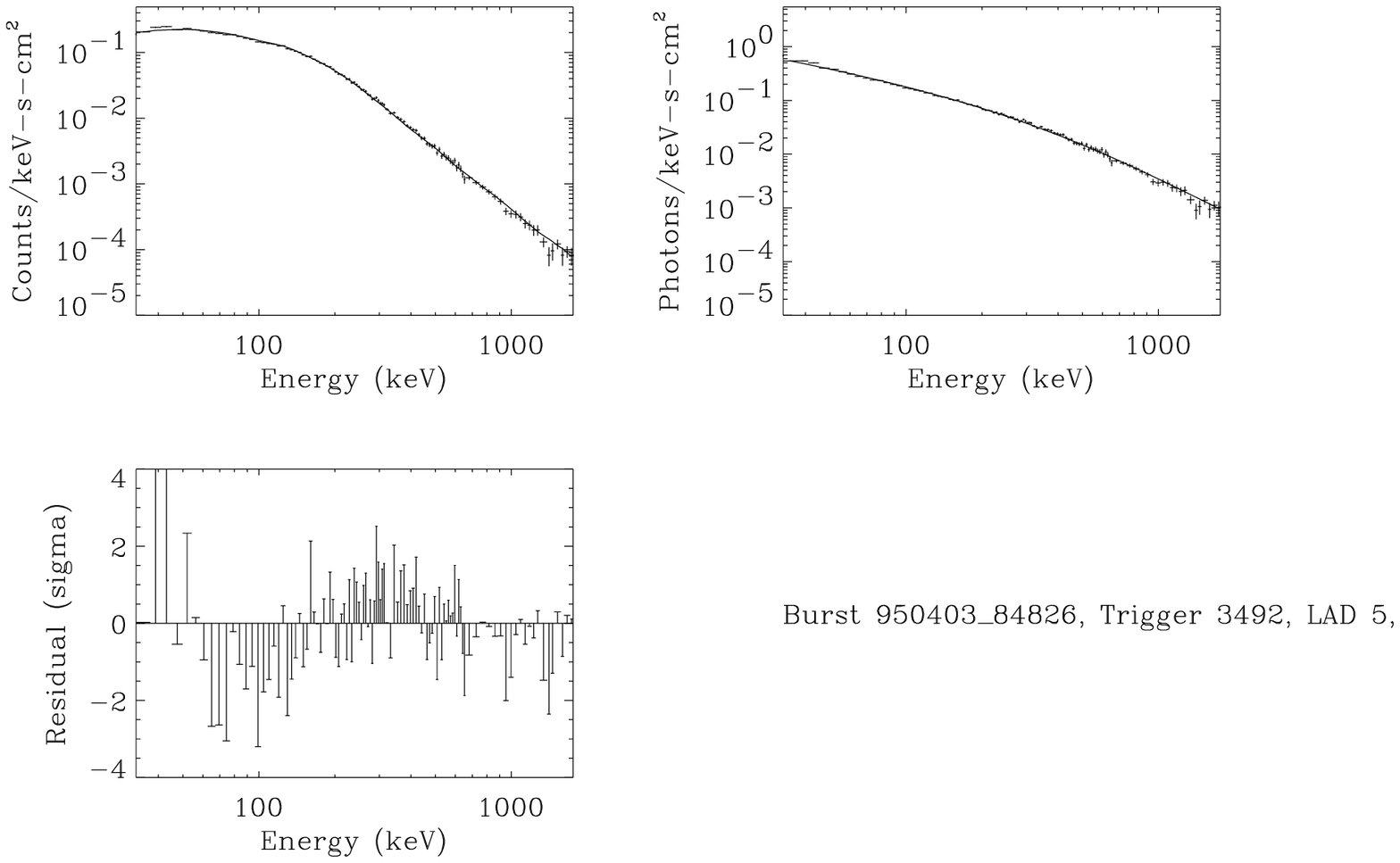,bbllx=116pt,bblly=578pt,bburx=295
pt,bbury=704pt,clip=true}}
\centerline{\psfig{file=comp.ps,bbllx=62pt,bblly=368pt,bburx=295
pt,bbury=524pt,clip=true}\psfig{file=smp.ps,bbllx=116pt,bblly=368pt,bburx=295
pt,bbury=524pt,clip=true}}
      \caption[!]{Trigger 3492. Spectral fits to the pulse time averaged spectrum. 
      The spectrum is integrated over the time interval [3.008-6.976] sec 
      since the trigger time. The model fits and the residuals are for the BAND, BPLW, 
      SSM and COMP model, displayed clockwise starting from the top left corner}
      \label{fig_5} \end{figure*}

In Table~\ref{averagefits} we report the best fit
parameters of the 4 spectral models obtained by fitting the
time averaged spectrum of the different peaks in each GRB.
For multi-peaked bursts the number of lines corresponds to
the number of peaks analyzed. There are some gaps in each
model columns corresponding to those pulses which are not
fitted (i.e. the model parameters are undetermined) by that
model. In Fig.~\ref{fig_5} is reported an example of fit:
the average pulse spectrum of trigger 3492 and the best fit
and residuals for each model are displayed.

\subsubsection{Comparison of the spectral models}

From the time integrated spectral analysis we do not find a
clear indication of a preferred fitting function to
represent the spectra. In fact all the models give
acceptable fits, and their $\chi^{2}_{red}$ are
around one for all the 4 models although their median is definitely grater than one. However we can report of an 
indication that the BAND model is the spectral shape which better
fits the time integrated spectra. 
The BAND model has an average $\chi^{2}_{red} \sim 1.3$, 
to be compared with the values 1.67, 1.63, 1.74 of the BPLW, 
COMP and SSM\footnote{The SSM model has to be considered however apart. 
In fact due to the considerable number of average peak spectra
which present an undetermined high energy supra-thermal
index ($\delta > 10$), the goodness of the model cannot
be easily quantified.} models,
respectively, and the width of the distributions, in terms
of standard deviations is 0.36 for the BAND and $\sim
0.68$ for the other three models. This indicates that the average pulse spectrum is better represented by the BAND model.
We performed a maximum likelihood test and obtained that the improvement of the
$\chi^{2}$ passing from the COMP model with 3 free parameters to the 4 free
parameter BAND model corresponds, in most cases, to a better fit, although there exists cases where they are statistically indistinguishable. 
This result agrees with what found by Band (\cite{Band b}).
Many exceptions exists anyway: as an example let just
mention the case of a spectrum whose high energy part is
too steep to be represented by the BAND model power law
(i.e. $\beta = -5$, first pulse of triggers 4368, 5567)
while the COMP model can accommodate this fast decreasing
spectrum with its exponential cutoff. Notice that in these
cases also the BPLW model gives acceptable fits for $\beta$: 
this is probably due to the sharp structure of this model 
which makes the fit to overestimate the high energy spectral 
hardness (i.e. less negative $\beta$ values).

Note also that a statistically good $\chi^{2}_{red}$ of
course cannot be considered a sufficient condition for the
quality of the fit and an analysis of the distribution of
the residuals is necessary.


\subsubsection{Spectral parameters distributions
and average spectral shape}

We first computed for each model the distributions of the
best fit parameters and found that they agree with the
results presented in previous spectral studies (Band et al.
\cite{Band b}). In particular for the BAND model we find
that there is no correlation between the low and the high 
energy spectral indices (the Pearson's correlation 
coefficient is $-$0.006) but they group in the ranges 
$-1.8\le \alpha \le - 0.6$ and $-3.5\le \beta \le -1.5$ 
which correspond to the intervals reported
by Band et al. (\cite{Band b}) and include the average values $<\alpha>=-1.03$ 
and $<\beta>=-3.31$ found by Fenimore (\cite{Fenimore}).  
The peak energy of the BAND model $E_{peak}\sim 205\pm 3\ \mathrm{keV}$ 
is instead harder than $<E_{peak}>=150\pm$50 keV reported in Band et al. 
(\cite{Band b})\footnote{The error associated with $<E_{peak}>$ represents the spread of the values and the uncertainty on $E_{peak}$ is the error on the average.}. This is probably due to the fact that our burst sample
represents the bright end of the BATSE peak flux
distribution, and because we integrated the time resolved
spectra just around the peak, excluding the major decaying
and rising parts of each pulse (see
sect.~\ref{sez:spectralana}) which are characterized
by softer spectra.

Table~\ref{aveave} reports the weighted average values of the best
fit parameters of Table~\ref{averagefits} for the pulse time 
integrated spectra.  

Secondly,  given that the analysis does not provide us with a
best fitting function, let us consider the robustness of the
parameters found with respect to the choice of the spectral model.

\noindent $\bullet$ The $\alpha$ distribution for the BAND model
is peaked around $-1$ and it is similar to the same
parameter distribution for the COMP model; the BPLW, due to
its sharp spectral break corresponding to the slope change
$E^{\alpha}\ \rightarrow \ E^{\beta}$, gives systematically
lower $\alpha$ values and, in fact, its average is $\simeq -1.2$.

\noindent $\bullet$ The $\beta$ distribution clusters around
$-2.3$ in the BAND and BPLW model and for the latter the
distribution is shifted towards higher values (i.e. harder
spectra). The same happens with the SSM model having an harder average 
high energy spectral tail ($\beta \sim -2.06$) then the BAND model.

\noindent $\bullet$ For the peak energy distribution we have
that the weighted average value for the BAND model  is $E_{peak} \sim 205\pm 3$ keV.
The BPLW break energy ($\equiv$ peak energy) 
is $\sim$169 keV, whereas the COMP model, due to the lack of the 
high energy power law component, overestimates the spectral break,
having a broad distribution and a weighted average of 233 keV. 
The SSM model peak energy is comparable with that of the BPLW model but lower than the BAND and COMP models.

We conclude that the average spectral shape of the GRBs present in our 
sample does depend on the fitting model and the BAND and COMP model tend to give,
although the latter lacks the high energy 
power law, comparable average spectral shape at low energies.

 \begin{table*} \caption[]{The peak average spectral results: for each
      trigger (col. 1) the best fit parameters, of the
      four models described in the text, are reported: for multi peaked 
      bursts all the peaks have their set
      of spectral parameters. The pulses which
      violate the low energy synchrotron limit are reported in Italics. Energy is in keV.}
\label{averagefits}
      \[ { \tiny \begin{array}{ccccccccccccccc} \hline
      \noalign{\smallskip} Trigger & \multicolumn{4}{c}{BAND} &
      \multicolumn{4}{c}{BPLW} & \multicolumn{3}{c}{COMP} &
      \multicolumn{3}{c}{SSM} \\ & $$\alpha$$ & $$\beta$$ & $$E_{0}$$
      $$^{\mathrm{a}}$$ & $$\chi^{2}_{r}$$ & $$\alpha$$ & $$\beta$$ &
      $$E_{0}$$ $$^{\mathrm{a}}$$ & $$\chi^{2}_{r}$$ &
      $$\alpha$$ & $$E_{0}$$ $$^{\mathrm{a}}$$  & $$\chi^{2}_{r}$$
      & $$E_{0}$$ $$^{\mathrm{a}}$$ & $$\delta$$
      &$$\chi^{2}_{r}$$ \\
            \noalign{\smallskip}
                \hline
            \noalign{\smallskip}

   143 & \textit{-0.59}$$\pm$$\textit{0.02} & \textit{-2.16}$$\pm$$\textit{0.09} & \textit{569}$$\pm$$\textit{33} & \textit{1.8} & -0.79$$\pm$$0.01 & -1.88$$\pm$$0.03 &
335$$\pm$$10 & 1.4 & -0.67$$\pm$$0.01 & 736$$\pm$$25 & 2.1
& 1208$$\pm$$136 & 7.2$$\pm$$1.8 & 3.2 \\
       & &  &  &  & -0.89$$\pm$$0.01 & -2.09$$\pm$$0.04 & 423$$\pm$$14 & 1.1 & -0.72$$\pm$$0.01 & 755$$\pm$$25 & 1.5 &
1184$$\pm$$191 & 10. & 2.22 \\
       & -0.79$$\pm$$0.02 & -3.21$$\pm$$0.45 & 568$$\pm$$24 & 1.6 & -1.00$$\pm$$0.01 & -2.24$$\pm$$0.04 & 358$$\pm$$10 & 1.1 & -0.80$$\pm$$0.01 & 590$$\pm$$17 & 1.6 &
728$$\pm$$64  & 8.8$$\pm$$2.3 & 2.0 \\
       & -1.37$$\pm$$0.03 & -2.49$$\pm$$0.25 & 677$$\pm$$79 & 1.5 & -1.52$$\pm$$0.01 & -2.27$$\pm$$0.06 & 281$$\pm$$18 & 1.4 & -1.4$$\pm$$0.02 & 784$$\pm$$61 & 1.5 &
86 $$\pm$$5  & 3.0$$\pm$$0.06 & 2.4 \smallskip \\
   1473 & -0.62$$\pm$$0.04 & -2.2 $$\pm$$0.09 & 326$$\pm$$27  & 1.2  & -1.15$$\pm$$0.02 & -1.89$$\pm$$0.01 & 137$$\pm$$6  & 1.1 & -1.09$$\pm$$0.01 & 467$$\pm$$25
& 1.3 & 600 $$\pm$$88  & 8.2$$\pm$$3.5 & 1.5 \\
        & -0.58$$\pm$$0.04 & -2.2 $$\pm$$0.06 & 280$$\pm$$22  & 1.6  & -0.94$$\pm$$0.02 & -1.96$$\pm$$0.03 &
210$$\pm$$7  & 2.2  & -0.78$$\pm$$0.02 & 444$$\pm$$20 & 1.9
& 553 $$\pm$$68  & 8.6$$\pm$$3.1 & 1.9 \\
        & -0.63$$\pm$$0.03 & -3.2 $$\pm$$0.16 & 183$$\pm$$6   & 1.7  & -1.15$$\pm$$0.01 & -2.41$$\pm$$0.03 &
181$$\pm$$2.8 & 4.6 & -0.68$$\pm$$0.02 & 198$$\pm$$5.0 &
1.9 &   &             &  \smallskip \\
    1541 & -0.9$$\pm$$0.04 & -1.98$$\pm$$0.04 & 248$$\pm$$24  & 1.0  &  &  &  &   &  -0.76$$\pm$$0.03 & 460$$\pm$$24
& 1.4 & 150$$\pm$$10  & 3.4$$\pm$$0.1 & 1.0 \\
   1625 & -0.88$$\pm$$0.02 & -2.07$$\pm$$0.07 & 608$$\pm$$42  & 0.9  & -1.06$$\pm$$0.01 & -1.90$$\pm$$0.03 & 300$$\pm$$11 & 1.1  &  -0.96$$\pm$$0.01 & 873$$\pm$$35 & 1.3 &
422 $$\pm$$24  & 3.7$$\pm$$0.2 & 1.0 \\
        & \textit{-0.58}$$\pm$$\textit{0.02} & \textit{-2.79}$$\pm$$\textit{0.15} & \textit{343}$$\pm$$\textit{15}  & \textit{1.4}  & -0.89$$\pm$$0.01 & -2.28$$\pm$$0.04 &
283$$\pm$$7. & 1.7  & -0.64$$\pm$$0.02 & 390$$\pm$$12 & 1.6
& 680 $$\pm$$82  & 10           & 3.2 \\
        & -0.71$$\pm$$0.02 & -3.75$$\pm$$0.6  & 256$$\pm$$10 & 1.2  & -1.09$$\pm$$0.01 & -2.62$$\pm$$0.05 & 244$$\pm$$5.7 & 1.7  &  -0.72$$\pm$$0.02 & 262$$\pm$$8.8 & 1.2 &
392 $$\pm$$39 & 10            & 2.6 \smallskip \\
   2083 & -0.80$$\pm$$0.01 & -2.71$$\pm$$0.07 & 261$$\pm$$7.4 & 3.1  &  &  &  &   &  -0.87$$\pm$$0.01 & 309$$\pm$$6.3& 3.8 &
343 $$\pm$$18  & 10.           & 3.2 \\
        & -0.71$$\pm$$0.04 & -3.4 $$\pm$$0.12 & 78 $$\pm$$3.  & 2.2  & -1.44$$\pm$$0.02 & -2.94$$\pm$$0.04 & 106$$\pm$$1.8& 3.2  &  -0.81$$\pm$$0.03 & 87 $$\pm$$2  & 2.3 &
106 $$\pm$$6.5 & 10.           & 4.8 \smallskip \\
   2329 & -1.09$$\pm$$0.01 & -1.78$$\pm$$0.03 &1136$$\pm$$65  & 2.6  & -1.17$$\pm$$0.01 & -1.69$$\pm$$0.01 & 311$$\pm$$10 & 2.2  &   &           &  &
175 $$\pm$$4  & 2.3$$\pm$$0.02& 6.0  \smallskip \\
   2537 & -1.05$$\pm$$0.05 & -2.7 $$\pm$$0.08 & 135$$\pm$$11  & 1.9  & -1.55$$\pm$$0.02 & -2.57$$\pm$$0.05 & 130$$\pm$$5  & 2.3  &  -1.25$$\pm$$0.03 & 193$$\pm$$10 & 2.5 &
84  $$\pm$$4  & 5. $$\pm$$0.2 & 1.8 \\
         & -1.08$$\pm$$0.06 & -2.69$$\pm$$0.06 & 109$$\pm$$9   & 2.9  & -1.62$$\pm$$0.02 & -2.58$$\pm$$0.04 & 108$$\pm$$3. & 3.6  &  -1.33$$\pm$$0.03 & 169$$\pm$$8.7& 3.6 &
65  $$\pm$$3  & 5.0$$\pm$$0.2 & 2.8 \\
        & -1.17$$\pm$$0.05 & -2.85$$\pm$$0.06 & 95 $$\pm$$7   & 4.3  &  &  &   &   &  -1.4 $$\pm$$0.03 & 137$$\pm$$6. & 4.7 &
53  $$\pm$$2.  & 5.4$$\pm$$0.2 & 3.9  \smallskip \\
   2798 & -0.86$$\pm$$0.01 & -2.4 $$\pm$$0.06 & 507$$\pm$$15  & 2.0  & -1.07$$\pm$$0.01 & -2.07$$\pm$$0.01 & 291$$\pm$$5  & 2.8  &  -0.9 $$\pm$$0.01 & 607$$\pm$$12 & 2.5 &
513 $$\pm$$20  & 6.0$$\pm$$0.5 & 2.3 \smallskip \\
   2831 & -0.54$$\pm$$0.04 & -5.0 & 870$$\pm$$112  & 1.6  & -0.70$$\pm$$0.02 & -1.88$$\pm$$0.05 & 451$$\pm$$23 & 1.7  &  -0.52$$\pm$$0.03 & 804$$\pm$$41 & 1.6 &
1919$$\pm$$577 & 7.$$\pm$$5. & 2.9 \\
         & -0.49$$\pm$$0.03 & -3.25$$\pm$$1.60 & 660$$\pm$$55  & 1.6  & -0.72$$\pm$$0.02 & -2.08$$\pm$$0.06 & 458$$\pm$$19 & 1.4  &  -0.50$$\pm$$0.03 & 676$$\pm$$32 & 1.6 &
1786$$\pm$$605 & 10.           & 3.2 \\
        &  &  &   &   & -0.93$$\pm$$0.02 & -1.96$$\pm$$0.06  & 412$$\pm$$25 & 1.18  &  -0.80$$\pm$$0.02 & 900$$\pm$$57 & 1.3 &
1033$$\pm$$208 & 6.7$$\pm$$3.1 & 1.4  \smallskip \\
   3481 & \textit{-0.59}$$\pm$$\textit{0.06} & \textit{-2.54}$$\pm$$\textit{0.20} & \textit{209}$$\pm$$\textit{20}  & \textit{1.3}  & -1.03$$\pm$$0.02 & -2.28$$\pm$$0.08 &
200$$\pm$$10 & 1.8  & -0.67$$\pm$$0.04 & 249$$\pm$$17 & 1.4
& 391 $$\pm$$91  & 10            & 1.6 \\
        & -0.85$$\pm$$0.02 & -2.1 $$\pm$$0.04 & 337$$\pm$$18  & 1.6  & -1.11$$\pm$$0.01 & -1.99$$\pm$$0.02 & 209$$\pm$$5.6& 2.4  &  -0.97$$\pm$$0.01 & 511$$\pm$$18 & 2.4 &
276 $$\pm$$12  & 4.0$$\pm$$0.2 & 1.5 \\
        & -1.09$$\pm$$0.04 & -2.55$$\pm$$0.12 & 194$$\pm$$16  & 1.9  & -1.33$$\pm$$0.03 & -2.13$$\pm$$0.03 & 104$$\pm$$4  & 2.4  &  -1.16$$\pm$$0.03 & 236$$\pm$$14 & 1.9 &
101 $$\pm$$6   & 4.4$$\pm$$0.2 & 1.6 \smallskip \\
   3491 & -0.92$$\pm$$0.05 & -2.11$$\pm$$0.09 & 298$$\pm$$37  & 1.1  & -1.19$$\pm$$0.02 & -1.98$$\pm$$0.04 & 175$$\pm$$9  & 1.2  &  -1.04$$\pm$$0.03 & 450$$\pm$$37 & 1.2 &
201 $$\pm$$19  & 3.8$$\pm$$0.3 & 1.1 \\
        & -1.09$$\pm$$0.05 & -2.2 $$\pm$$0.12 & 394$$\pm$$57  & 1.1  & -1.29$$\pm$$0.02 & -2.03$$\pm$$0.05 & 186$$\pm$$13 & 1.0  &  -1.19$$\pm$$0.03 & 574$$\pm$$54 & 1.2 &
147 $$\pm$$13  & 3.4$$\pm$$0.2 & 1.2 \\
        & -0.93$$\pm$$0.04 & -2.6 $$\pm$$0.09 & 197$$\pm$$12  & 1.3  & -1.28$$\pm$$0.02 & -2.30$$\pm$$0.03 & 145$$\pm$$4  & 1.5  &  -1.04$$\pm$$0.02 & 252$$\pm$$11 & 1.6 &
165 $$\pm$$9 & 6.7$$\pm$$0.4 & 1.2 \smallskip  \\
 3492 & -1.05$$\pm$$0.02 & -2.38$$\pm$$0.07 & 546$$\pm$$27  & 1.3  & -1.25$$\pm$$0.01 & -2.15$$\pm$$0.02 & 292$$\pm$$8  & 1.5  &  -1.11$$\pm$$0.01 & 702$$\pm$$21 & 1.8 &
243 $$\pm$$8   & 3.7$$\pm$$0.1 & 2.0 \smallskip \\
 3523 &                  &                  &               &      & -0.96$$\pm$$0.01 & -1.77$$\pm$$0.01 & 315$$\pm$$6.8& 1.5  &  -0.87$$\pm$$0.01 &1003$$\pm$$24 & 2.6 &
682 $$\pm$$31  & 3.9$$\pm$$0.2 & 1.7 \smallskip  \\
 4368 & -1.76$$\pm$$0.02 & -5.              & 813$$\pm$$113 & 1.1  & -1.87$$\pm$$0.01 & -2.42$$\pm$$0.04 & 218$$\pm$$15 & 1.2  &  -1.78$$\pm$$0.02 & 782$$\pm$$60 & 1.0 &
31  $$\pm$$1.4 & 3.3$$\pm$$0.03& 1.7 \\
      & -1.73$$\pm$$0.03 & -2.9 $$\pm$$0.16 & 357$$\pm$$30 & 1.4  & -1.91$$\pm$$0.01 & -2.58$$\pm$$0.04 & 155$$\pm$$7  & 1.5  &  -1.76$$\pm$$0.02 & 410$$\pm$$25 & 1.3 &
37. $$\pm$$1.  & 3.9$$\pm$$0.05& 1.7 \smallskip \\
   5477 &                  &                  &               &      & -0.79$$\pm$$0.03 & -1.38$$\pm$$0.04 & 288$$\pm$$31 & 0.9  &  -0.78$$\pm$$0.03 & 1765$$\pm$$192& 1.0 &
937 $$\pm$$208 & 2.3$$\pm$$0.3 & 0.9 \\
        &                  &                  &               &      & -0.87$$\pm$$0.01 & -1.66$$\pm$$0.07 & 602$$\pm$$56 & 1.3  &  -0.80$$\pm$$0.02 & 1765$$\pm$$120& 1.8 &
925 $$\pm$$126 & 2.3$$\pm$$0.2 & 1.6 \smallskip \\
   5563 & -1.00$$\pm$$0.07 & -2.53$$\pm$$0.11 & 169$$\pm$$20  & 1.2  & -1.42$$\pm$$0.03 & -2.36$$\pm$$0.05 & 140$$\pm$$7 & 1.2  &  -1.15$$\pm$$0.04 & 236$$\pm$$17 & 1.3&
118 $$\pm$$9.7 & 4.9$$\pm$$0.4 & 1.2 \smallskip \\
   5567 & -1.93$$\pm$$0.11 & -5.0             & 442$$\pm$$244 & 1.2  & -2.10$$\pm$$0.06 & -2.42$$\pm$$0.17 & 114$$\pm$$34 & 1.2   &  -1.93$$\pm$$0.10 & 436$$\pm$$195& 1.1 &
25$$\pm$$5   & 3.7$$\pm$$0.2 & 1.1 \\
        & -1.58$$\pm$$0.07 & -2.26$$\pm$$0.36 & 686$$\pm$$282 & 1.2  & -1.69$$\pm$$0.03 & -2.28$$\pm$$0.17 & 219$$\pm$$40 & 1.1  &  -1.61$$\pm$$0.04 & 861$$\pm$$242& 1.2 &
35  $$\pm$$6   & 2.9$$\pm$$0.1 & 1.4 \\
        & -1.38$$\pm$$0.05 & -3.19$$\pm$$2.8  & 403$$\pm$$62  & 1.3  & -1.58$$\pm$$0.02 & -2.42$$\pm$$0.12 & 194$$\pm$$17 & 1.2  &  -1.39$$\pm$$0.04 & 413$$\pm$$52 & 1.3 &
74 $$\pm$$7   & 3.5$$\pm$$0.2 & 1.6 \\
         & -1.32$$\pm$$0.03 & -3.3 $$\pm$$1.2  & 570$$\pm$$58  & 1.6  & -1.5 $$\pm$$0.01 & -2.45$$\pm$$0.08 & 271$$\pm$$16 & 1.4  &  -1.32$$\pm$$0.02 & 584$$\pm$$47 & 1.6 &
102 $$\pm$$6.  & 3.3$$\pm$$0.1 & 2.8 \\
         & -1.13$$\pm$$0.02 & -2.8 $$\pm$$0.2  & 532$$\pm$$38  & 1.4  & -1.3 $$\pm$$0.01 & -2.18$$\pm$$0.04 & 242$$\pm$$10 & 1.3  &  -1.15$$\pm$$0.02 & 578$$\pm$$29 & 1.5 &
180 $$\pm$$9   & 3.7$$\pm$$0.1 & 2.2 \smallskip \\
   5614 &        &              &       &         &  -1.22$$\pm$$0.01 & -2.67$$\pm$$0.03 & 227$$\pm$$3  & 3.4  &  -0.9 $$\pm$$0.01 & 273$$\pm$$5. & 1.4 &
277 $$\pm$$10  & 10.0            \smallskip    \\
    5621 & -0.70$$\pm$$0.04 & -2.13$$\pm$$0.04 & 193$$\pm$$14  & 1.2  & -1.08$$\pm$$0.02 & -2.00$$\pm$$0.02 & 144$$\pm$$4. & 1.8  &  -1.02$$\pm$$0.02 & 400$$\pm$$17 & 2.4 &
213 $$\pm$$11  & 4.3$$\pm$$0.2 & 1.5 \\
         & -0.99$$\pm$$0.09 & -2.2 $$\pm$$0.06 & 156$$\pm$$26  & 1.2  & -1.19$$\pm$$0.07 & -2.05$$\pm$$0.03 & 86 $$\pm$$5. & 1.2  &  -1.30$$\pm$$0.04 & 323$$\pm$$32 & 1.5 &
90. $$\pm$$8.  & 3.9$$\pm$$0.2 & 1.1 \smallskip \\
   5704 & -1.21$$\pm$$0.07 & -2.7 $$\pm$$0.28 & 190$$\pm$$24  & 0.9  & -1.60$$\pm$$0.03 & -2.51$$\pm$$0.08 & 146$$\pm$$9  & 1.0  &  -1.26$$\pm$$0.05 & 215$$\pm$$19 & 0.9 &
85  $$\pm$$9   & 4.6$$\pm$$0.4 & 1.0 \smallskip \\
   6198 & -1.03$$\pm$$0.01 & -2.53$$\pm$$0.06 & 317$$\pm$$11  & 1.7  & -1.30$$\pm$$0.01 & -2.23$$\pm$$0.02 & 196$$\pm$$3. & 2.4  &  -1.1 $$\pm$$0.01 & 387$$\pm$$10.& 2.4 &
184 $$\pm$$5.  & 4.5$$\pm$$0.1 & 2.2 \smallskip \\
   6404 & \textit{-0.06}$$\pm$$\textit{0.2}  & \textit{-1.9} $$\pm$$\textit{0.05} & \textit{68} $$\pm$$\textit{15}  & \textit{1.0}  & -0.89$$\pm$$0.09 & -1.90$$\pm$$0.04 & 89.$$\pm$$5. & 1.1  &  -0.87$$\pm$$0.07 & 207$$\pm$$21 & 1.3 &
180 $$\pm$$35  & 5.5$$\pm$$1.7 & 1.2 \\
        & \textit{-0.19}$$\pm$$\textit{0.14} & \textit{-2.6} $$\pm$$\textit{0.12} & \textit{65} $$\pm$$\textit{8}   & \textit{1.1}  & -1.05$$\pm$$0.06 & -2.41$$\pm$$0.06 & 96 $$\pm$$4  & 1.5  &  -0.53$$\pm$$0.08 & 92 $$\pm$$6  & 1.4 &
159 $$\pm$$36  & 10.           & 2.3 \smallskip \\
   6581 & -1.43$$\pm$$0.08 & -2.07$$\pm$$0.29 & 549$$\pm$$209 & 0.8  & -1.55$$\pm$$0.04 & -1.92$$\pm$$0.07 & 146$$\pm$$23 & 0.8  &  -1.50$$\pm$$0.05 & 730$$\pm$$179& 0.8 &
51.6$$\pm$$8.  & 2.8$$\pm$$0.1 & 0.8 \smallskip \\
   7301 & -1.14$$\pm$$0.16 & -2.3 $$\pm$$0.1  & 127$$\pm$$31  & 0.7  & -1.6 $$\pm$$0.06 & -2.3 $$\pm$$0.06 & 96 $$\pm$$8  & 0.8  &  -1.4 $$\pm$$0.06 & 240$$\pm$$33 & 0.9 &
59  $$\pm$$7   & 4. $$\pm$$0.3 & 0.8 \\
         & -1.10$$\pm$$0.03 & -3.0 $$\pm$$0.4  & 372$$\pm$$28  & 1.1  & -1.37$$\pm$$0.01 & -2.45$$\pm$$0.07 & 250$$\pm$$11 & 1.5  &  -1.12$$\pm$$0.02 & 392$$\pm$$23 & 1.2 &
172 $$\pm$$11  & 4.2$$\pm$$0.3 & 1.4 \\
        & -0.89$$\pm$$0.02 & -2.3 $$\pm$$0.10 & 500$$\pm$$33  & 1.3  & -1.11$$\pm$$0.01 & -2.06$$\pm$$0.04 & 290$$\pm$$10 & 1.8  &  -0.95$$\pm$$0.02 & 621$$\pm$$26 & 1.5 &
398 $$\pm$$26  & 4.6$$\pm$$0.4 & 1.3 \\
       & -1.16$$\pm$$0.04 & -2.27$$\pm$$0.18 & 546$$\pm$$78  & 0.9  & -1.33$$\pm$$0.02 & -1.99$$\pm$$0.05 & 222$$\pm$$16 & 1.1  &  -1.2 $$\pm$$0.03 & 677$$\pm$$64 & 0.9 &
137 $$\pm$$11  & 3.0$$\pm$$0.1 & 0.9 \\
       & -1.17$$\pm$$0.04 & -1.87$$\pm$$0.1  & 724$$\pm$$153 & 0.8  & -1.3 $$\pm$$0.02 & -1.8 $$\pm$$0.05 & 230$$\pm$$23 & 0.8  &  -1.25$$\pm$$0.02 & 1184$$\pm$$153& 0.8 &
114 $$\pm$$11  & 2.5$$\pm$$0.1 & 0.9 \smallskip \\
  7549 &                  &                  &               &      & -1.19$$\pm$$0.01 & -1.76$$\pm$$0.02 & 247$$\pm$$11 & 1.5  &  -1.13$$\pm$$0.01 & 1192$$\pm$$59 & 1.4 &
182$$\pm$$8   & 2.6$$\pm$$0.05& 1.7 \\
       & -0.98$$\pm$$0.03 & -2.5 $$\pm$$0.3  & 494$$\pm$$46  & 1.2  & -1.19$$\pm$$0.01 & -2.17$$\pm$$0.07 & 282$$\pm$$15 & 1.4  & -1.01$$\pm$$0.02 & 558$$\pm$$36 & 1.2    &
276$$\pm$$22  & 4.0$$\pm$$0.3 & 1.3 \\
        & -0.68$$\pm$$0.02 & -2.64$$\pm$$0.18 & 397$$\pm$$23  & 1.1  & -0.94$$\pm$$0.01 & -2.07$$\pm$$0.04 & 262$$\pm$$8.3& 1.9  &  -0.72$$\pm$$0.02 & 446$$\pm$$18 & 1.2 &
650 $$\pm$$82  & 8.7$$\pm$$3.2 & 1.7 \\
       & -0.97$$\pm$$0.03 & -2.2 $$\pm$$0.07 & 277$$\pm$$22  & 1.2  & -1.28$$\pm$$0.01 & -2.08$$\pm$$0.04 & 180$$\pm$$7  & 1.8  &  -1.09$$\pm$$0.02 & 396$$\pm$$21 & 1.6 &
161 $$\pm$$9   & 3.8$$\pm$$0.2 & 1.1 \\
       & -0.84$$\pm$$0.02 & -2.67$$\pm$$0.09 & 201$$\pm$$8   & 2.0  & -1.22$$\pm$$0.01 & -2.24$$\pm$$0.03 & 156$$\pm$$3.1& 4.3  &  -0.91$$\pm$$0.02 & 232$$\pm$$7  & 2.3 &
231 $$\pm$$15  & 8.4$$\pm$$1.6 & 1.7 \\
        \noalign{\smallskip}
        \hline
    \end{array}
     }  \]
\begin{list}{}{}
\item[$^{\mathrm{a}}$] Energy in keV.
\end{list}
   \end{table*}

\subsubsection{The Synchrotron limit violation}

The analysis of the low energy spectral indices shows that
there are 4 pulses (those in italics in
Table~\ref{averagefits}) whose average BAND best fit
spectrum is harder, at $2\sigma_{\alpha}$ level, than the
limit $E^{-2/3}$ predicted by the synchrotron model. Three of 
these spectra, when fitted with the COMP model, show no $\alpha$
limit violation but their low energy power law spectral
index ($-0.667\pm 0.01$, $-0.64\pm 0.02$ and $-0.53\pm 0.08$) is 
very close to $-2/3$.
The same spectra have a poor fit with the SSM model showing 
systematic trend in the 
residuals of the fits on most of the energy range. 
The average pulse spectra give only a weak indication of the violation of
the synchrotron model low energy spectral index.
On the other hand, as will be shown in the following, stronger
indications of a violation come from the time resolved spectra.
This occurs because the averaging of the spectral 
evolution over the duration of the pulse systematically yields
softer spectra. 

   \begin{table}
      \caption[]{Weighted averages of the averaged--peak spectral 
      parameters for the four models.  
      The break energy $\mathrm{E_{break}}$ and the peak energy
      $\mathrm{E_{peak}}$ are in keV.}
         \label{aveave}
      \[ {\scriptsize
         \begin{array}{ccccc}
            \hline
            \noalign{\smallskip}
                         &  BAND   &  BPLW & COMP & SSM$$^{\mathrm{a}}$$ \\
            \noalign{\smallskip}
            \hline
            \noalign{\smallskip}
            <$$\alpha$$>  &  -0.977$$\pm$$ 0.003 & -1.185$$\pm$$0.002 & -1.013$$\pm$$0.002 &   -2/3 (fix)        \\
            <$$\beta$$>   & -2.27$$\pm$$0.01     &  -2.09$$\pm$$0.004 &                    & -2.06$$\pm$$0.06    \\
            <$$\mathrm{E_{break}}$$> & 164.4$$\pm$$1.7 &  161$$\pm$$0.7  & 210.4$$\pm$$1.3 &  67.4$$\pm$$0.7     \\
            <$$\mathrm{E_{peak}}$$> & 205$$\pm$$3 &  161$$\pm$$0.7  & 233$$\pm$$2 &  168$$\pm$$2           \\
            \noalign{\smallskip}
            \hline
         \end{array}   }
      \]
\begin{list}{}{}
\item[$^{\mathrm{a}}$] For this model the spectral parameters are
described in the text. Notice that $\beta$ represents the
high energy power law spectral index derived from the slope 
$\delta$ of the particle distribution, as $\beta=-(\delta+1)/2$.
\end{list}
   \end{table}

\subsection{Time resolved spectra}

Let us now consider the results of the fitting of the time
resolved spectra with the same four models.

\begin{figure}
    \resizebox{\hsize}{10cm}{\includegraphics{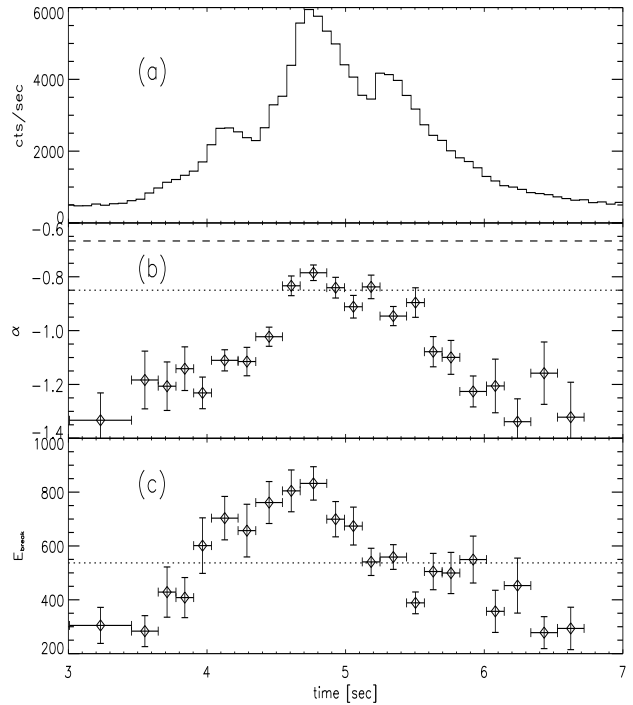}}
      \caption[]{Trigger 3492. Spectral evolution of the COMP model
      parameters fitted to the time resolved spectra. Light curve on the 64ms
      time-scale (panel a), for the energy range 110--320 keV (corresponding to       channel 3 of the detector); low energy spectral index (b), \textit{dotted      line}:
      the average $\alpha$ value from Table.\ref{averagefits}, \textit{dashed line}:
      the synchrotron model limit; peak energy (c), \textit{dotted line}:
      the average $E_{break}$ of the COMP model. } \label{fig_11}
\end{figure}

In Fig.~\ref{fig_11} an example of spectral
evolution is reported: for each peak we obtain a sequence of best fit
parameters relative to the time resolved spectra which
characterize the temporal evolution of the spectrum during the burst. 
While a full description and discussion on the
parameter evolution will be presented elsewhere (Ghirlanda
et al, in preparation), we note that both the peak energy
and the low energy spectral index evolve in phase: the
flux and the spectrum becomes harder during the rise phase
and softer during the decay, a behavior already found by 
Ford et al. (\cite{Ford a}) as a characteristic 
spectral evolution morphology of several bursts.

In the following we report and comment on the parameter
distributions of the 4 models for a comparison (i) among 
the model themselves and (ii) with the results on the pulse 
average spectrum.
Finally we examine the SSM limit violation in some
well defined cases.

\subsubsection{Comparison of the spectral models}

For a general comparison on the quality of the fits with
the different models we have plotted in Fig.~\ref{fig6}
their reduced $\chi^{2}$ distributions. 
These are all centered around one and again it is not possible 
to identify any preferable spectral model even considering their spread.
Anyway we can note some differences: considering that all these 
distributions are asymmetric towards 2, we fitted a 6 parameter 
function, namely a right asymmetric gaussian distribution, and obtain 
that the BAND and COMP model have the lower dispersed distributions with 
$\sigma \sim 0.14$, to be compared with $\sim 0.2$ for the BPLW and SSM model.
This result indicates that in terms of reduced $\chi^{2}$
the BAND and COMP model could better represent the time
resolved spectra of bright bursts.
Also in this case some counter examples
exist showing that in general within a
single pulse time resolved spectra can be fitted by
different spectral models (see Fig.~\ref{fig_12}).
We also tested, with a Kolmogorov--Smirnov 
test, whether the BAND and COMP distributions could have been drawn from the
same distribution and obtain that this is the case with a probability of 
0.99 (95\% confidence level).

\subsubsection{Time resolved vs time integrated spectra}

\begin{figure*}
\centerline{\psfig{file=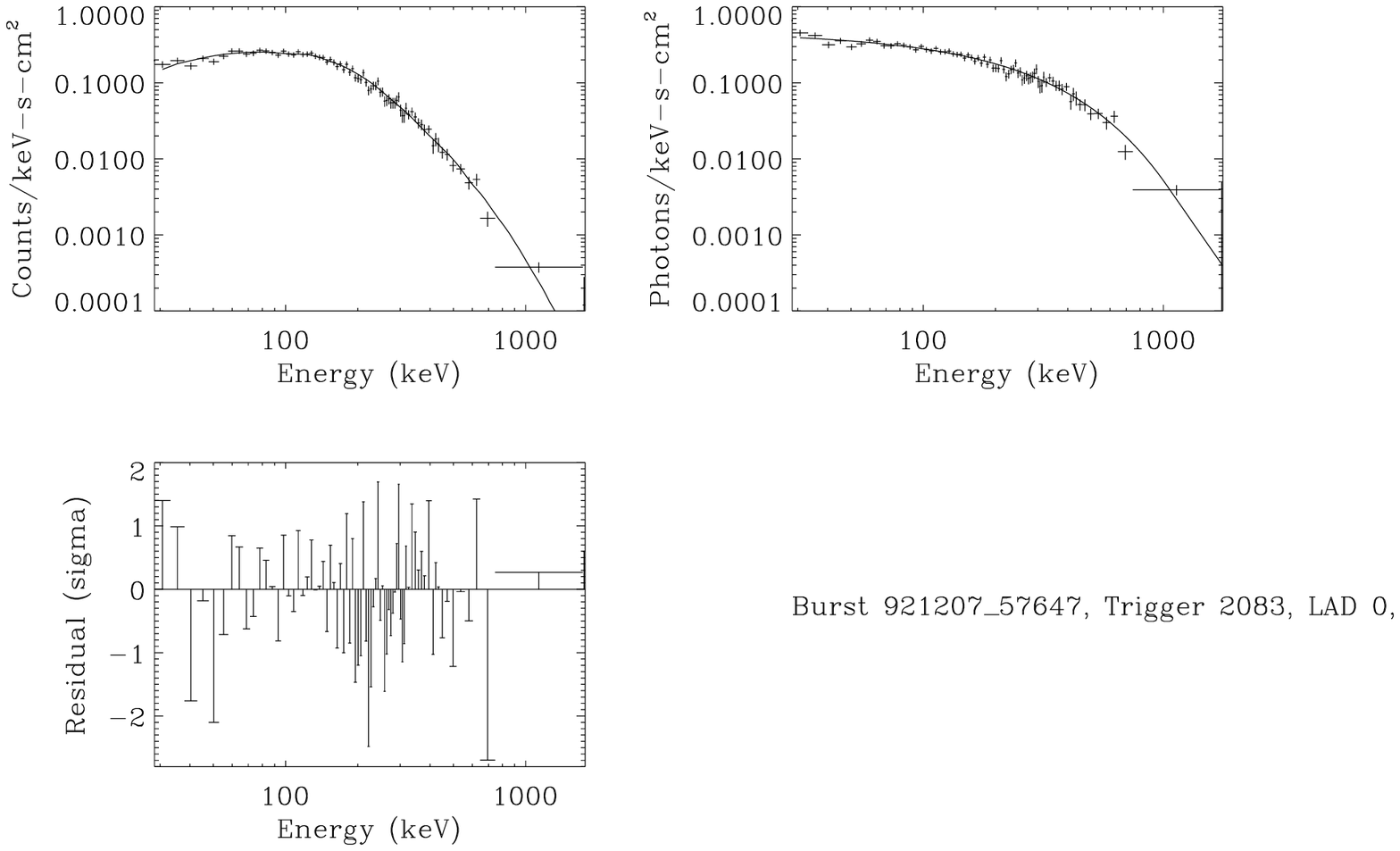,bbllx=60pt,bblly=578pt,bburx=295
pt,bbury=704pt,clip=true}\psfig{file=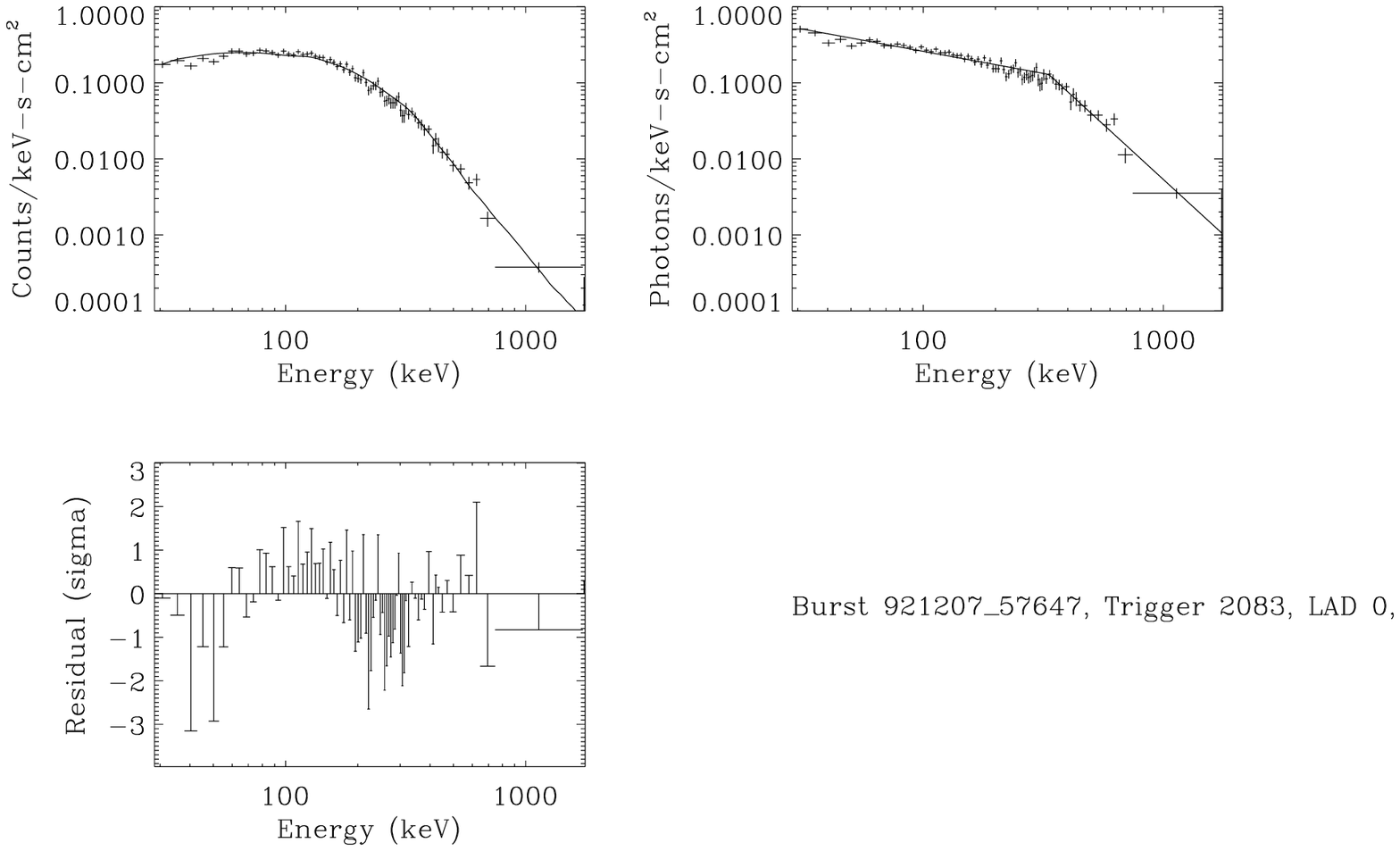,bbllx=116pt,bblly=578pt,bburx=295
pt,bbury=704pt,clip=true}}
\centerline{\psfig{file=grb_peak.ps,bbllx=60pt,bblly=368pt,bburx=295
pt,bbury=524pt,clip=true}\psfig{file=bplw_peak.ps,bbllx=116pt,bblly=368pt,bburx=295
pt,bbury=524pt,clip=true}}
\centerline{\psfig{file=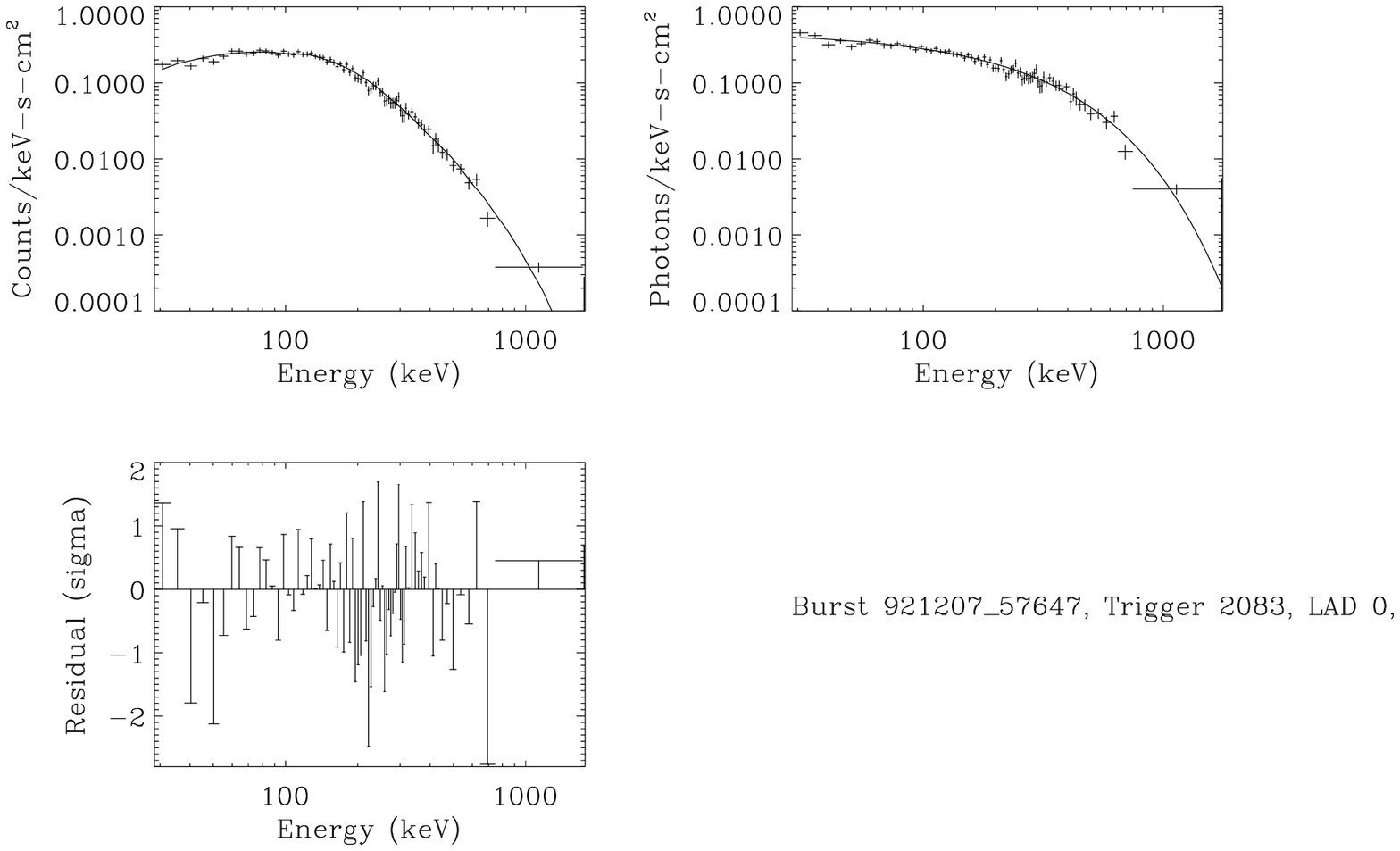,bbllx=60pt,bblly=578pt,bburx=295
pt,bbury=704pt,clip=true}\psfig{file=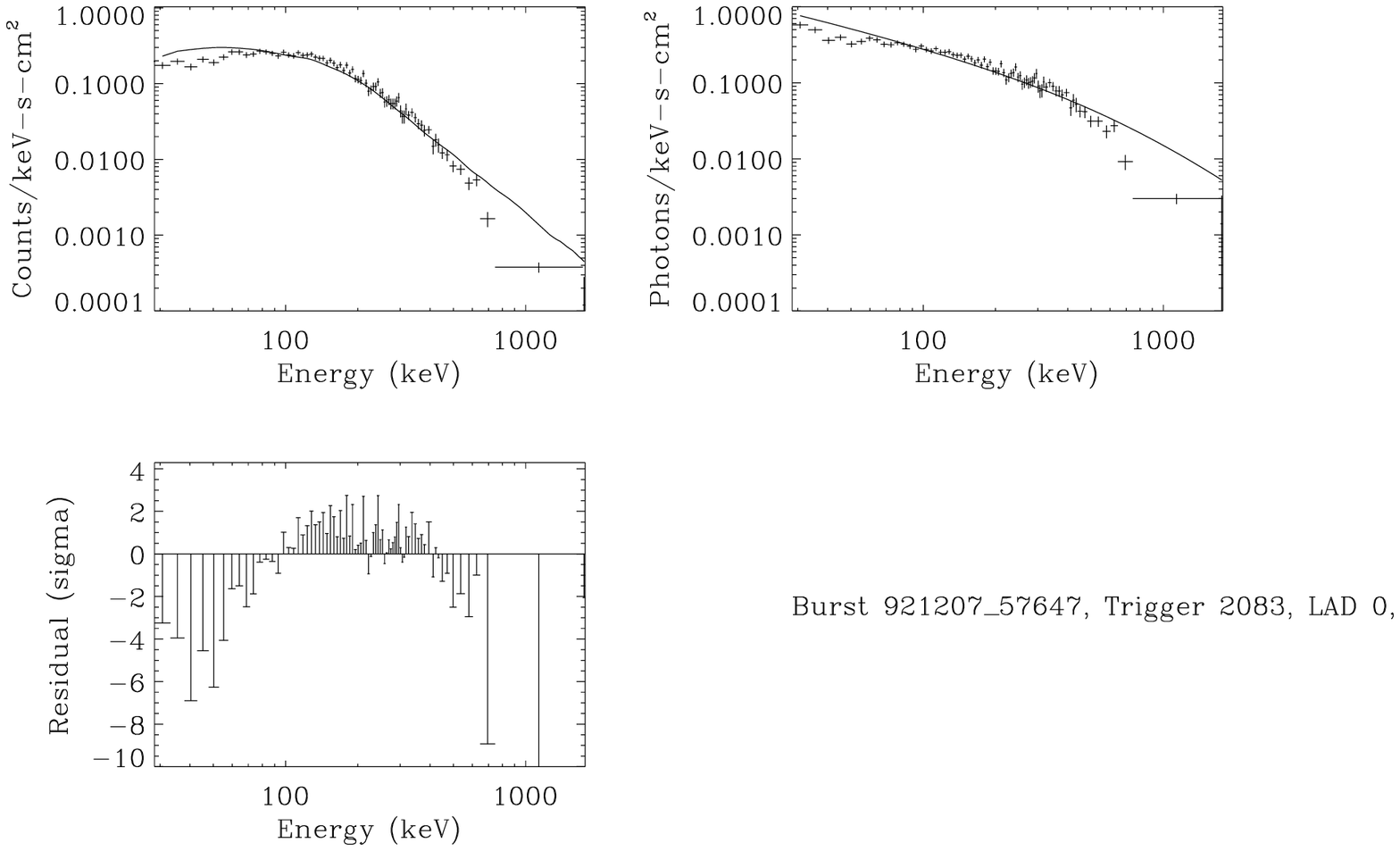,bbllx=116pt,bblly=578pt,bburx=295
pt,bbury=704pt,clip=true}}
\centerline{\psfig{file=comp_peak.ps,bbllx=60pt,bblly=368pt,bburx=295
pt,bbury=524pt,clip=true}\psfig{file=smp_peak.ps,bbllx=116pt,bblly=368pt,bburx=295
pt,bbury=524pt,clip=true}}
      \caption[!]{Trigger 2083. Spectral fits to the pulse peak
      spectrum. The spectrum is integrated over the time interval
      [1.088-1.216] sec since the trigger time. The model fits and the
      residuals are for the BAND, BPLW, SSM and COMP model displayed
      clockwise starting from the top left corner.} \label{fig_12}
      \end{figure*}

\begin{figure}
    \resizebox{\hsize}{!}{\includegraphics{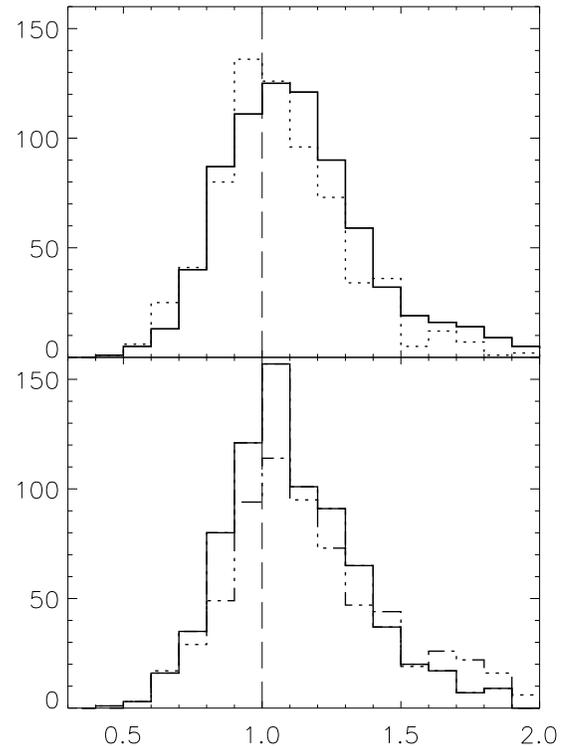}}
      \caption[]{Reduced $\chi^{2}$ distributions for the 4 spectral
      models. The total number of time resolved spectral fits is 700
      for the BAND and BPLW (\textit{dotted and solid line} top panel) and  
      COMP and SSM (\textit{solid and dot--dashed line} bottom panel).}
      \label{fig6}
\end{figure}

Let us now compare with the corresponding results for the
time integrated spectra. 
As an example in Fig.~\ref{fig_12}
we report the peak spectrum of the trigger 2083. Comparing with
Fig~\ref{fig_5} (showing the spectrum time integrated over 
the whole peak, for trigger 4392) 
it is evident that the time resolved spectra 
better constraint the best fitting model: 
in fact (for example in this
case) the time averaged pulse spectrum was satisfactorily
fitted by all the 4 models whereas the peak time resolved
spectrum is not well fitted by the SSM model 
because the low energy powerlaw spectral index is fixed 
at -2/3 and the spectrum is harder than this slope 
($\alpha=-0.03\pm 0.06$ for the BAND model). 
In this example the BAND, COMP and BPLW model provide the best 
fits and among them the BAND model has the best $\chi^{2}_{red}=0.99$.

This represents a clear indication that the time
resolved spectra have to be used in determining the
spectral properties of GRBs: average spectra can be
effective in comparing the global properties of different
bursts, but the actual spectral shape requires high temporal
resolution data.

%

\subsubsection{Spectral parameters distributions}

In this Section we consider the distribution
of the model parameters as inferred from the time resolved
data, compare again the different models,and compare these results
with the time integrated ones and with previous findings.

\begin{itemize}

\item {\it The low energy spectral component.}

The BAND, COMP and BPLW fits are comparable at low energies
as shown by the corresponding $\alpha$ distributions presented in
fig.~\ref{fig7}: the BAND and COMP model distributions
are similar, as in the case of the time integrated spectra, 
and both have a mode of $-0.85\pm0.1$, consistent at $2\sigma$ with the 
BPLW average value $-1.15 \pm 0.1$. Note that qualitatively the
extension of the $\alpha$ distribution of the BPLW model
towards lower values could be attributed to the fact that
at low energies the sharp break tends to
underestimate the hardness of the spectrum compared to a
smoothly curved model.

The average low energy spectral slope obtained from the time 
resolved spectra is harder than what obtained with the time 
integrated pulse spectra for all the three models (BAND, BPLW, COMP). 
This is a consequence of time integration 
(i.e. hardness averaging) of the spectral evolution 
(which can be also very dramatic) 
over the entire rise and/or decay phase of the pulse.

\begin{figure}
\resizebox{\hsize}{!}{\includegraphics{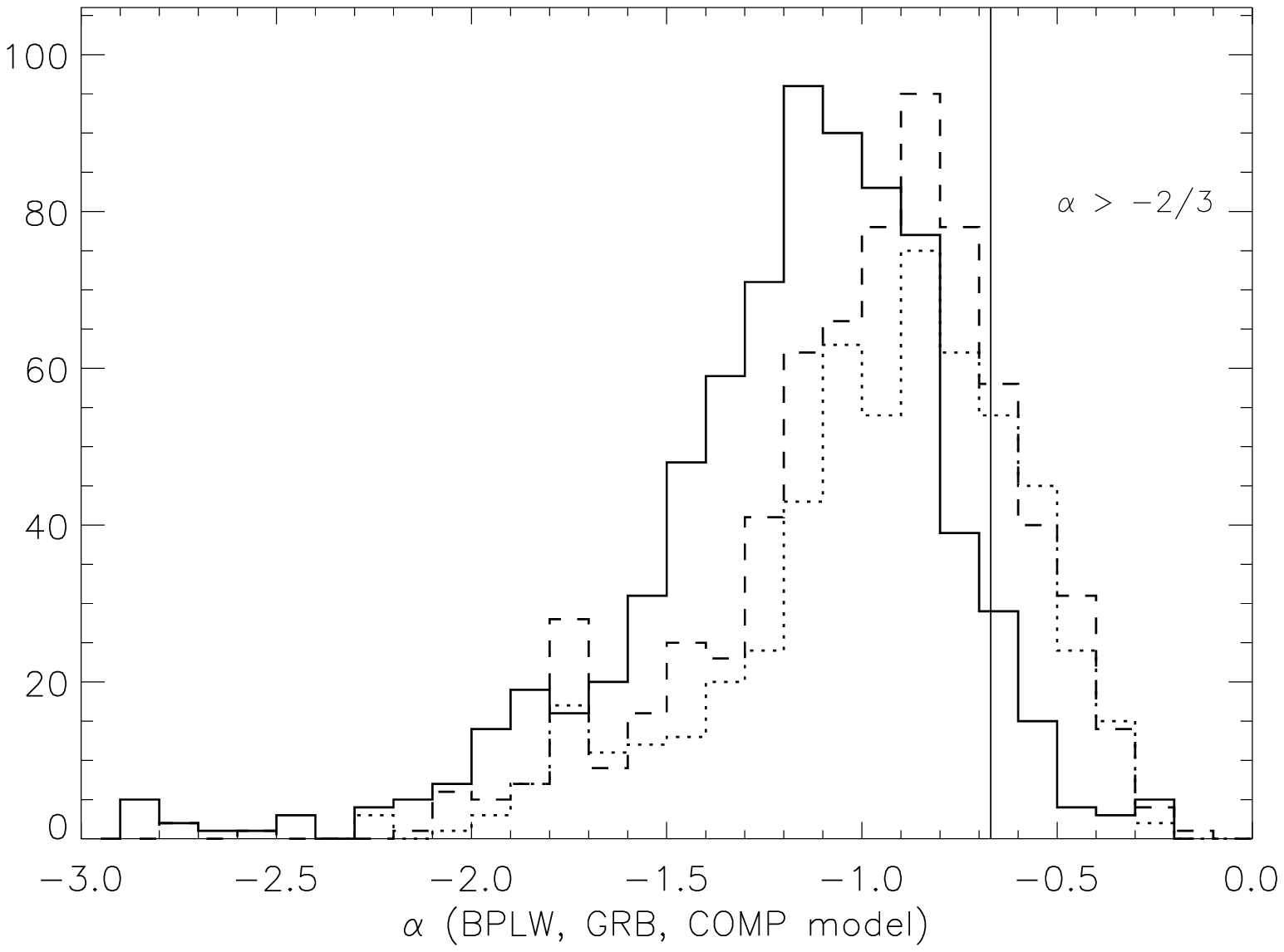}}
      \caption[]{Low energy power law spectral index ($\alpha$)
      distributions derived from the time resolved spectral
      analysis. \textit{Solid line:} BPLW model, \textit{dotted line:}
      BAND model, \textit{dashed line:} COMP model. The vertical line
      represents the synchrotron limit ($\alpha=-2/3$) for the low
      energy spectral shape.} \label{fig7}
\end{figure}

Even though we present the spectral analysis with 4 models
separately we can compare these results with those obtained
by Preece et al. (\cite{Preece2000}), where they considered
the low energy spectral index distribution regardless of
the (best) fitted model. We obtain a similar distribution but
with a harder average low energy spectrum
($\alpha \sim -0.85\pm 0.1$ considering the BAND and COMP
model, to be compared with $\sim -1.05 \pm 0.1$). This difference
might be a consequence of the fact that we are considering
a subsample of that of Preece et al. (\cite{Preece2000}),
including only the bright GRBs which might be intrinsically
characterized by a harder spectrum (Borgonovo \& Ryde
\cite{Borgonovo}) and because we restricted our time
resolved spectral analysis to the peak interval excluding
the inter-pulse phase of multi-peaked events.

\vskip 0.3 true cm

\item {\it The high energy spectral component.}

In many spectra the steepness of the count distribution
above the break energy $E_0$, coupled with the poor S/N
ratio, result in a poorly determined value of $\beta$. This
is true especially in the case of the BAND model: when the
break energy $E_0$ sets at high values the exponential
roll--over leaves too few high energy spectral channels
free for fitting the $E^{\beta}$ component properly.
In Fig.~\ref{fig8} we report the BPLW, BAND and SSM 
$\beta$ distributions. Also in this case the BPLW model,
due to its sharpness, tends to
overestimate the hardness of the count spectrum giving
systematically higher values of $\beta$ than the BAND
model. The mode of this parameter is $-2.45 \pm 0.1$ and
$-2.05 \pm 0.1$ for the BAND and BPLW model, respectively.

The SSM average $\beta$ is $-2.17$ which is consistent with 
what found from the average pulse spectral analysis.
The average $\beta$ value for the BAND model ($-2.45\pm 0.1$) 
is softer than the time averaged results, and the BPLW model
value ($-2.05$) is consistent with the results of Table.~\ref{aveave}.

In fig.~\ref{fig8} it is also indicated the critical value  $-2$: 
spectra with $\beta \ge -2$ are rising in $E F_E$, and we can 
set only a lower limit to $E_{peak}$. These cases
correspond to the 18\% and 7\% of the spectra fitted with
the BAND and BPLW model and the 16\% for the SSM. 
Notice that there is also a subclass of soft-spectra with 
$\beta \le -3$ which are characterized by a very steep 
spectral tail: these spectra are clearly better fitted 
by the COMP model.

The spread in these $\beta$ distributions ($-4 \le \beta \le -1.5$) 
corresponds to what found by Band et al. (\cite{Band b}).

\begin{figure}
    \resizebox{\hsize}{!}{\includegraphics{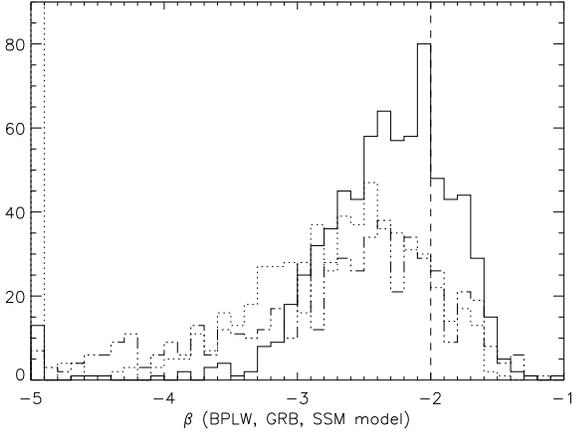}}
      \caption[]{High energy power law spectral index ($\beta$)
distributions for the time resolved spectra. \textit{Solid
line:} BPLW model; \textit{dotted line:} BAND model; \textit{dot--dashed line:} SSM model,
 in this case $\beta$ is calculated from the $\delta$ parameter (see text). 
Also shown (bin with $\beta$=-5) the time resolved spectra with undetermined 
high energy spectral index for the BAND model. The dashed vertical line indicates the critical value $\beta = -2$.}
\label{fig8}
\end{figure}

\vskip 0.3 true cm
\item {\it The spectral break.}

The most important spectral parameter obtained in fitting
the spectrum with these models is $E_{peak}$.
As just mentioned this
characteristic energy can be obviously calculated only for
those spectra (BPLW and BAND model) with $\beta < -2$.

\begin{figure}     
    \resizebox{\hsize}{!}{\includegraphics{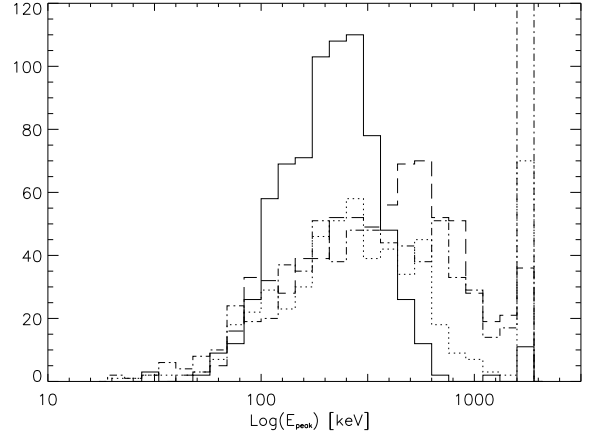}}
      \caption[]{Peak energy distribution for the 4 spectral models. 
\textit{solid line:} BPLW model, \textit{dotted line:} BAND model, 
\textit{dashed line:} COMP model, \textit{dot--dashed line:}  SSM model. 
Spectra with undetermined 
peak energy (i.e. the high energy threshold 1800 keV assumed as lower limit) 
are reported in the last bin.}
      \label{fig9}
\end{figure}

In Fig.\ref{fig9} the peak energy distributions for the
various models are reported and it is also shown the bin with 
$E_{peak}$=1800 keV, assumed as lower limit of the peak energy
for those spectra in which the BAND, BPLW or SSM model have $\beta \ge -2$.

The mode is $E_{peak} = 280^{+72}_{-57}$ keV for the BAND model,
consistent, within its error bar, with the BPLW most
probable value of $211^{+25}_{-22}$ keV. The COMP model,
instead, gives a highly asymmetric peak energy distribution
with a mode of $595^{+104}_{-88}$ keV because the lack of
a high energy power law component tends to overestimate
the energy corresponding to the start of the exponential
cutoff.
The SSM model has an average $E_{peak} \sim 316^{+64}_{-52}$ keV
with a wide distribution.

From the analysis of the average spectral shape, obtained from the
time integrated and from the time resolved
spectral analysis, we would like to stress that if we want
to characterize the spectral hardness it is
necessary to extend the single parameter analysis,
typically based on the peak energy or the hardness ratio,
and consider the low and high energy spectral components.

\end{itemize}

\begin{figure}
    \resizebox{\hsize}{10cm}{\includegraphics{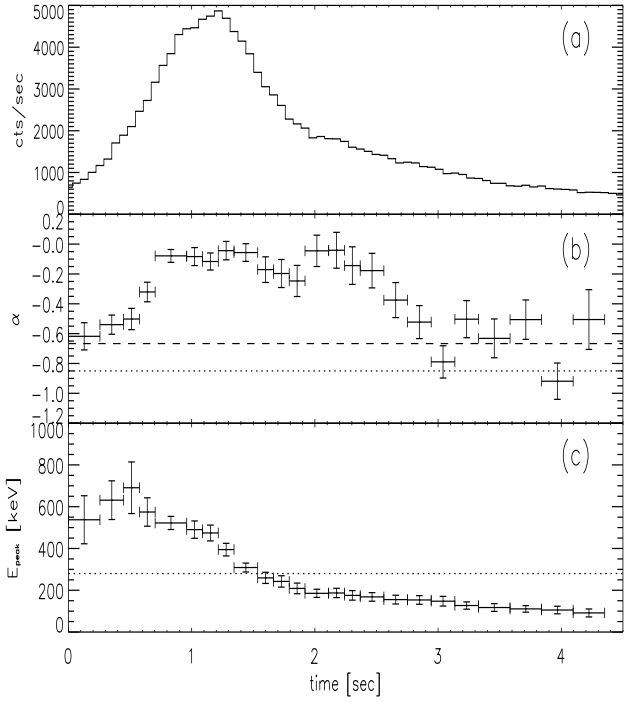}}
      \caption[]{Trigger 2083. Spectral evolution of the BAND model
      parameters fitted to the time resolved spectra. 
      Light curve on the 64ms time-scale (panel a) for the energy range 110--320 keV (corresponding to       channel 3 of the detector); 
      low energy spectral index and (\textit{dashed}) synchrotron 
      shock model limit $\alpha=-2/3$ (b); peak energy (c). 
      For reference the average values of $\alpha$ and $E_{peak}$ 
      obtained from the time resolved spectra (\textit{dotted line}) 
      and the synchrotron model limit (\textit{dashed line}) are reported. } 
\label{fig11}
\end{figure}

\subsection{The Synchrotron limit violation}

A well known prediction of the optically thin
synchrotron model is that the asymptotic low energy photon
slope $\alpha$ should be lower than or equal to $-2/3$ (Katz
et al. \cite{Katz}).  From the analysis of the time
resolved spectra we obtain that not only the low energy
power law slope can violate this limit, but it can also
evolve dramatically
during a single pulse 
(as already found by Crider et al. \cite{Crider a},b).

\begin{figure}
    \resizebox{\hsize}{10cm}{\includegraphics{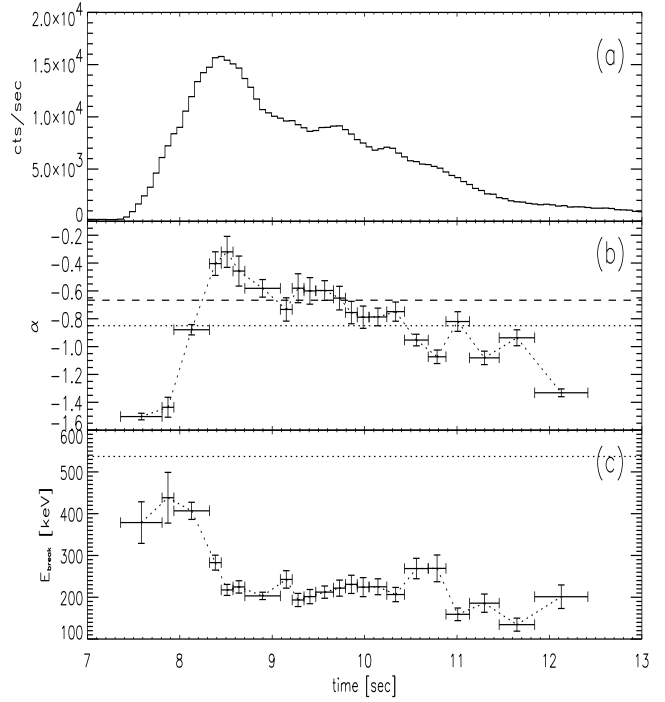}}
      \caption[]{Trigger 5614. Spectral evolution of the COMP model
      fitted to the time resolved spectra. Light curve on the 64ms
      time-scale panel (a) for the energy range 110--320 keV (corresponding to       channel 3 of the detector); low energy spectral index and
      (\textit{dashed}) synchrotron shock model limit $\alpha=-2/3$
      (b); peak energy (c).
      For reference are reported the average
      values of $\alpha$ and $E_{peak}$ obtained from the time resolved spectra
      (\textit{dotted line}) and the synchrotron model limit 
      (\textit{dashed line}).
      }\label{fig12}
\end{figure}

We can characterize this behaviour, for example, via the BAND and
COMP model fits (we exclude the BPLW model which, as showed
above, tends to underestimate the hardness of the spectrum
at low energies). We obtain that the 14\% of the time
resolved spectra fitted with the BAND model are
inconsistent with $\alpha \le -2/3$ at 2$\sigma$.
A similar percentage ($\sim$11.7\% of course mostly for 
the same spectra) of spectra violating the
$\alpha$ limit is found for the COMP model.

For these extremely hard spectra no correlation between
$\alpha$ and any other fit parameter is found, and the
violation occurs both in the rise and decay phase of the
pulses. Moreover $\sim$ 21\% of the time resolved
spectra corresponding to the peak time bin violate at $2\sigma$ the synchrotron limit, 
indicating that this violation happens during the peak 
phase and not preferentially before or after it. As an example in
Fig.~\ref{fig11} we show the spectral evolution of $\alpha$ 
in the case of trigger GRB921207 (BAND model). During the main peak
(t$\in$[0.0,4.5] sec) the majority of the time resolved
spectra violates the synchrotron model limit ({\it dashed
line} in panel (c)). Notice that during this time interval
the low energy spectral index evolves between $-2/3$ and
0.0 and the peak energy (panel (b)) decreases
monotonically. Another interesting example is reported in
Fig.~\ref{fig12}. GRB960924 shows a low energy spectral
shape harder than $-2/3$ only around the peak, and the
$\alpha$ evolution during the rise phase covers an interval
of $\Delta \alpha \sim 1.4$.

The predictions of the synchrotron model have been recently 
discussed by Lloyd \& Petrosian (\cite{Lloyd}, L\&P2001). They claim 
that the low energy spectral limit $-2/3$ is not so constraining 
if (among other assumptions) an anisotropic pitch angle distribution 
is assumed for the emitting electrons (which should be the case 
in a low density and intense magnetic field regime). 
In the case of small pitch angles ($\Psi \ll 1$) the emitted spectrum 
should be characterized by three components: a low energy flat count 
spectrum $\nu^{0}$ (for $\nu \ll \nu_{s} \equiv 2/3\ \nu_{b}/(\gamma_{m}\ 
\Psi^{2})$) followed by the typical $\nu^{-2/3}$ power law 
(for $\nu_{s} \le \nu \le \nu_{m}$) and then by $\nu^{-(p+1)/2}$. 
\textit{p} is the electron power law energy index, $\nu_{s}$ the 
break at the transition from the flat spectrum to the $-2/3$ slope, 
and $\nu_{m}$ represents the frequency corresponding to $\gamma_{m}$ 
where the electron energy distribution is smoothly cut off (at low energies).

According to L\&P2001 $\alpha \ge -2/3$ is allowed and the slope 
$\nu^{\alpha}$ obtained from fitting a two component model (like the BAND 
function) to such a three components spectrum is typically an average 
between $\nu^{0}$ and $\nu^{-2/3}$. 
During  the emission  the  electrons cool  so  that their  low
energy limit $\gamma_{m}$ decreases and  as a consequence the peak energy
becomes softer $E_{peak}\propto \gamma_{m}^{2}$. The electrons average
pitch angle decreases (although their distribution can be still anisotropic)
and   this  causes   the  frequency   $\nu_{s}   \propto  (\gamma_{m}\
\Psi^{2})^{-1}$ to increase. The  spectral evolution predicted by this
model then is  of hardening of the low energy  powerlaw -- because of the
progressive disappearance of the $\nu^{-2/3}$ spectral component -- while
the peak energy  naturally evolves from hard to  soft.  Thus in burst
violating the  synchrotron limit,  a negative correlation  between the
peak energy and the low energy spectral index is expected.

\begin{figure}
    \resizebox{\hsize}{10cm}{\includegraphics{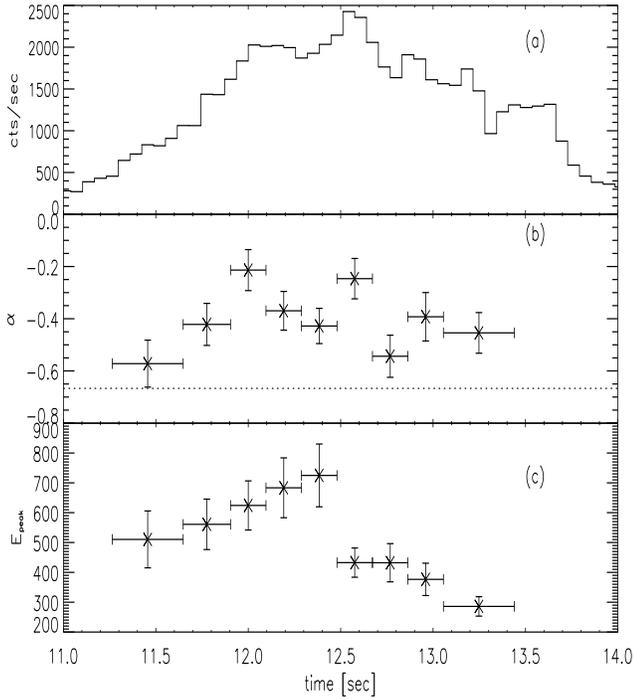}}
      \caption[]{Trigger 1625. Spectral evolution of the BAND model
      parameters fitted to the time resolved spectra. Light curve on the 64ms
      time--scale (panel a) for the energy range 110--320 keV (corresponding to       channel 3 of the detector); low energy spectral index and
      (\textit{dashed}) synchrotron shock model limit $\alpha=-2/3$
      (b); peak energy (c).}\label{fig13}
\end{figure}

\begin{figure}
    \resizebox{\hsize}{10cm}{\includegraphics{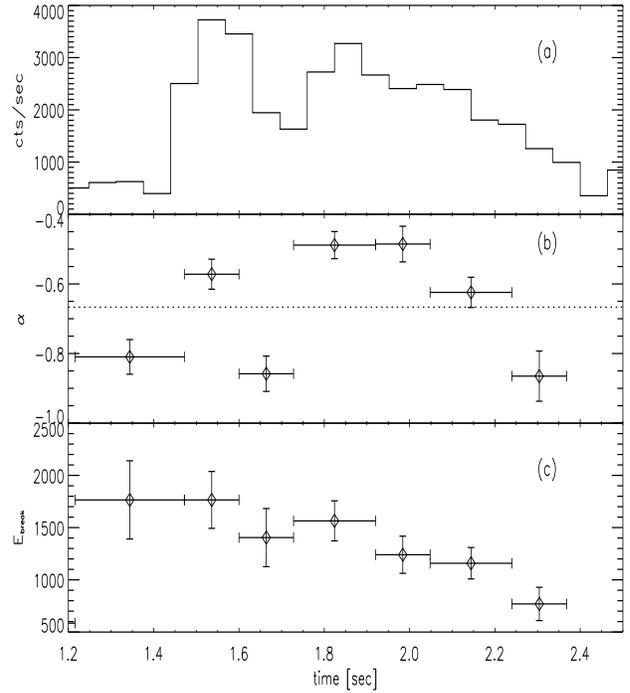}}
      \caption[]{Trigger 5477. Spectral evolution of the BAND model
      parameters fitted to the time resolved spectra. Light curve on the 64ms
      time-scale (panel a) for the energy range 110--320 keV (corresponding to       channel 3 of the detector); low energy spectral index and
      (\textit{dashed}) synchrotron shock model limit $\alpha=-2/3$
      (b); peak energy (c).  } \label{fig14}
\end{figure}

We report in Fig.~\ref{fig13} as an example the evolution during GRB 920525
which shows a low energy spectral component harder than $-2/3$ during the 
main peak (for $\sim$ 2 sec). We note (panel (b) and (c) in Fig.~\ref{fig13}) 
that there is no evidence for negative correlation between the peak energy 
and $\alpha$: during the rise phase of the flux (panel (a)) the peak energy 
increases regardless of the rise and partial decay of the spectral index, while
for the rest of the pulse the peak energy decreases and alpha varies above 
the $-2/3$ limit. 
The same happens for GRB 921207 (Fig.~\ref{fig11}) and GRB 960529 
(Fig.~\ref{fig14}): in these bursts the peak energy decreases while 
$\alpha$ goes above and below the synchrotron limit. 

\section{Conclusions}

We considered a sample of bright bursts detected by BATSE
and performed a uniform analysis for the time integrated
and the time resolved (typically 128 ms) spectra with four
different models proposed in the literature.

We find that even with this time resolution no parametric
model can better represent the data and different spectra require
different shapes, re--confirming the erratic behaviour of
bursts and also possibly indicating that time resolution on
time--scales comparable with the variability one is needed to shed light
on such erratic characteristics.

Indeed, an important result we confirm is that the
average time integrated spectrum often used in the
literature does not well represent the very same event
resolved on shorter time-scales.  The time integrated
spectra might still be used for a comparison of the
average spectral shape among different pulses and as 
indicators of the average spectral
parameters of the time resolved analysis although only 
the time resolved spectra should be used in any test of a 
physical emission model.

Finally, a considerable number of the fitted spectra are
characterized by extremely hard low energy components with
spectral index $\alpha$ greater than $-2/3$, value
predicted by synchrotron theory (Katz \cite{Katz}).
This violation was found by Crider et al. (\cite{Crider a}) and has been 
recently reported by Frontera et al. (\cite{Frontera}) in some GRBs
observed by BeppoSAX. They report 1 sec time resolved spectra 
significantly harder than $E^{-2/3}$, mainly during the first 
phase of the burst emission.
We have found that in 11 of the 25 bursts analyzed the $\alpha$ limit 
violation is evident mainly in the spectra around the peak both during 
the rise and decay phase, and this could indicate that at least at some 
stages of the burst evolution -- possibly near the peak of emission
itself -- radiative processes, other than synchrotron, can dominate 
the emission.

We also reported some examples of bursts which violate the synchrotron 
limit and are not characterized by the $\alpha$--$E_{peak}$ 
anticorrelation predicted by the small pitch angle distribution 
synchrotron model proposed by Lloyd \& Petrosian (\cite{Lloyd}).

The obvious extension of this work, namely the study and
discussion of the temporal evolution of the spectral shape,
will be the subject of a forthcoming paper (Ghirlanda et
al., in preparation).


\begin{acknowledgements}
This research has made use of data obtained through the
High Energy Astrophysics Science Archive Research Center
Online Service, provided by the NASA/Goddard Space Flight
Center.  We are grateful to D. Band for useful
discussions and suggestions for the BATSE data analysis and then for his careful reading of the manuscript and useful comments on it as referee. 
We are also grateful to M. Tavani for providing his code for the
SSM model. Giancarlo Ghirlanda and AC
acknowledge the Italian MIUR for financial support.
\end{acknowledgements}

\end{document}